\documentclass[a4paper,twocolumn,11pt,floatfix,amssymb]{quantumarticle}
\pdfoutput=1

\usepackage{dsfont,amsfonts,mathtools,amsmath,amsthm}    
\usepackage{graphicx}   
\usepackage{verbatim}   
\usepackage{subfigure}
\usepackage{hyperref} 
\usepackage[ngerman,english]{babel}
\usepackage[usenames, dvipsnames]{xcolor}
\usepackage{tikz}
\usepackage{enumitem}
\usetikzlibrary{plotmarks,arrows.meta,calc,decorations.markings,math,arrows.meta,decorations.pathreplacing,decorations.pathmorphing,patterns}
\usepackage[percent]{overpic}
\def\be{\begin{equation}}
\def\ee{\end{equation}}
\DeclareMathOperator{\Tr}{Tr}
\DeclareMathOperator*{\Ltri}{\text{\raisebox{0.25ex}{\scalebox{1.}{$\blacktriangleleft$}}}}
\DeclareMathOperator*{\Rtri}{\text{\raisebox{0.25ex}{\scalebox{1.}{$\blacktriangleright$}}}}
\usepackage{braket}

\begin{document}

\title{Topological Phase Transitions Induced by Varying Topology and Boundaries in the Toric Code}

\author{Amit Jamadagni}
\email{amit.jamadagni@itp.uni-hannover.de}
\affiliation{Institut f\"ur Theoretische Physik, Leibniz Universit\"at Hannover, Appelstra{\ss}e 2, 30167 Hannover, Germany.}
\author{Arpan Bhattacharyya}
\email{abhattacharyya@iitgn.ac.in}
\affiliation{Indian Institute of Technology, Gandhinagar, Gujarat 382355, India}

\begin{abstract}
One of the important characteristics of topological phases of matter is the topology of the underlying manifold on which
they are defined. In this paper, we present the sensitivity of such phases of matter to the underlying topology, by
studying the phase transitions induced due to the change in the boundary conditions. We claim that these phase transitions
are accompanied by broken symmetries in the excitation space and to gain further insight we analyze various
signatures like the ground state degeneracy, topological entanglement entropy while introducing the open-loop operator
whose expectation value effectively captures the phase transition. Further, we extend the analysis to an
open quantum setup by defining effective collapse operators, the dynamics of which cool the system to distinct steady
states both of which are topologically ordered. We show that the phase transition between such steady states is effectively captured by the
expectation value of the open-loop operator.
\end{abstract}

\maketitle

\section{Introduction}

Topological phases are phases of matter whose description is beyond the Landau symmetry breaking theory. Due to the absence of a local order parameter, it is challenging to detect and classify such phases of matter. Several signatures such as Ground State Degeneracy (GSD), Topological Entanglement Entropy (TEE) \cite{Jiang2012}, modular S and U matrices
\cite{Morampudi2014} have been effective in detecting a Quantum Phase Transition (QPT) between Topologically Ordered (TO) and trivial phases. On similar lines, there has been recent interest in detecting a QPT between two distinct topological phases,
termed as Topological Phase Transition (TPT) \cite{Morampudi2014, Zarei2016, Hu2018, Iqbal2018}. We investigate the presence of a TPT based on the notion of Hamiltonian deformation as in Ref.~\cite{Castelnovo_2010}. We consider a TPT induced by a parameterized
Hamiltonian, $H(\lambda)$, which at the extremities of the parameter reduce to a frustration-free Hamiltonian. In such scenarios, the presence of a TPT is signalled by the energy gap closing or the change in the GSD as we interpolate between the endpoints \cite{Yoshida2011}. 

Topological phases of matter with intrinsic topological order have been well understood in models with periodic boundary
conditions \cite{Kitaev2003, Levin2005} while the systematic classification of open boundaries has been gaining significance in
the recent times \cite{Beigi2011, Kitaev2012, Cong2017}. It has a twofold purpose. It, not only helps us to gain an insight into different topological phases of matter, thereby
providing a means to classify different phases \cite{Iqbal2018}, but also open boundaries form a more natural setting in
experimentally realizing topological phases \cite{Fowler2012,Sameti2017}. In this paper, we aim to understand the
sensitivity of the topological phases of matter to different boundary conditions. To this extent, we analyze the presence of a
TPT by interpolating between different boundary variations of the Toric Code (TC) model. In Sec.~\ref{sec_ffh} we introduce the
TC Hamiltonian in a general setting, briefly motivating the different boundary conditions. We then provide necessary arguments
which consolidate the presence of a TPT, further we comment on the broken symmetries that accompany the TPT. In
Sec.~\ref{sec_ugsd}, we present various scenarios where the phase transitions are marked by the change in the GSD, while in
Sec.~\ref{sec_egsd}, we present scenarios where the phase transitions are captured by the closing of the energy gap at some
interpolation strength. In each of the above sections, we introduce phase transitions which are induced by varying the underlying
topology and by varying the open boundary conditions. For each of the transitions, we introduce an open-loop operator and claim
that its expectation value is sensitive to different phases and hence effectively captures the phase transition. 
\par While QPT's in closed systems have been extensively studied, the study of the same in an open quantum setting has gained
traction recently \cite{Weimer2015, Vincent2017, Meghana2018}. The understanding of these, on one hand, help in identifying and
classification of new phases of matter \cite{Helmrich2018, Carollo2019} while on the other hand help tune experimental setups
where external interaction is inevitable \cite{Weimer2010, Sameti2017}. Lastly, in Sec.~\ref{sec_idiss}, we sketch a procedure to
realize the TPT's of the closed system in an open quantum setup. We engineer dissipative collapse operators which effectively
cool the system to distinct steady states depending on the strength of the interpolation parameter. The effective cooling
rate of the collapse operators in the open system context is analogous to the interpolation strength of the closed system while
the steady states of the open system at the extremities of interpolation get mapped to the respective ground states of the closed
system. Using the fact that TPT in an open system is encoded in the properties of the steady-state, we show that the expectation
value of the open-loop operator is still effective in detecting such phase transitions.

\section{Connecting frustration-free Toric Code Hamiltonians \label{sec_ffh}}

We begin by briefly reviewing the general features of the TC model with different boundary conditions.
Consider a square lattice with vertices (faces) denoted by $v$($p$), with spins on the edges of the
lattice. The general TC Hamiltonian is given by
\begin{align}
\label{eq_tc}
H = -\sum_{v}A_{v} - \sum_{p}B_{p},
\end{align}
with $A_{v} = \prod_{i}\sigma_{x}^{(i)}$ and $B_{p}=\prod_{j}\sigma_{z}^{(j)}$ where $i(j)$ denote the spins attached
to the respective vertices (faces). For periodic boundary conditions, four spins are attached to each vertex (face) as
in Fig.~\ref{fig1}(b). The excitations in the system (also referred to as anyons) are given by $A_{v}$, $B_{p}$
violations, denoted by $e$, $m$ respectively and are generated by $\sigma_{z}$ and $\sigma_{x}$ operators.


As introduced in Ref.~\cite{Beigi2011}, we define the boundary as an interface between a TO phase and vacuum and
classify different boundaries by the behavior of the excitations at the boundary. At a given boundary, every excitation
either gets identified with vacuum and is called condensing excitation, or, is retained at the boundary and is called
non-condensing excitation. 
For the case of TC, we identify the boundary where  $e$($m$) excitations condense as rough (smooth)
boundary. For both the above mentioned cases, the Hamiltonian still retains the form of Eq.~\ref{eq_tc}, with
$A_{v}$, $B_{p}$ operators being modified at the boundary, for instance Eq.~\ref{eq5} at $\lambda=0$,
$\lambda=1$ represent the interaction at the rough and the smooth boundary. For a formal mathematical treatment of 
boundaries we refer the reader to Ref.~\cite{Beigi2011}.

Due to the different condensation properties at a given boundary, each boundary condition gives rise to a unique
topological phase. If they were to belong to the same phase it would immediately imply that there exists a local unitary
transformation connecting the ground states \cite{Chen2010}, further implying that the excitations belonging to different
sectors are unitarily equivalent. In other words, if the phase with periodic boundary conditions were to belong 
to the same phase as the open boundary, it would imply the existence of local unitary transformation connecting the 
ground states  of the above phases which would further imply that the excitations from both phases are related via 
the unitary. The above scenario is not possible, as otherwise it would imply the existence of non-trivial anyon 
condensation in the periodic boundary i.e., in the absence of a physical boundary. Similarly, we can extend
the above notion to conclude that phases with different physical boundaries are distinct as otherwise it would imply
the existence of local unitary transformation mapping a non-condensing excitation to a condensing excitation and vice-versa.
Additionally, the ground state of the toric code with periodic boundaries is given by a superposition 
of closed loops where as in the case of open boundaries the superposition includes open loops and therefore
the ground states with periodic and open boundaries conditions cannot be mapped via local unitaries. The 
above argument can also be extended in comparing the ground states of different open boundary conditions as 
the open loops appearing in the superposition are different due to different anyon condensation.
The difference in the structure of the superposition of loops in the ground states further consolidates the 
fact that different boundary conditions give rise to distinct topological phases and therefore, 
interpolating different boundary conditions via Hamiltonian interpolation encapsulates a TPT. 



To further consolidate the above notion of a TPT, we introduce the notion of parity conservation and anyonic symmetries.
We claim that the break in either one of the symmetries is sufficient to encode a TPT. It is well established that the
excitations in the TC model with periodic boundaries appear in pairs, with the introduction of boundary this 
parity is no longer conserved as it is possible to draw relevant single excitations from the boundary.
Another symmetry in the case of the TC is given by the fact that the fusion and braiding rules of
excitations remain invariant under the exchange of the labels $e \leftrightarrow m$, which is commonly referred to as
electric-magnetic duality/anyonic symmetry \cite{Bombin2010,Teo2016}. For the case of periodic boundary condition, the
anyonic symmetry is retained (upto the presence of a domain wall) while in the open boundary context the anyonic symmetry
is broken due to change in fusion rules at the boundary. We further note that, to encode a TPT it is sufficient that
either one of the symmetry is broken but it is not necessary that every TPT 
is accompanied by a broken symmetry. We further elaborate on the above statement in the appendix by providing a suitable
example, and also introduce additional constraints on the parity symmetry so as to complete the bi-implication.

We present different TPT's obtained by interpolating between different boundary conditions, i.e., by tuning the $A_{v}, B_{p}$
interactions to
\begin{enumerate}
    \item vary the underlying topology, i.e., breaking the periodicity with introduction of open boundaries (effective topology variation)
    \item vary the open boundary conditions, with the underlying topology intact (effective boundary variation)
\end{enumerate}
As the above variations encompass a variety of scenarios, we further classify the phase
transitions into the following two classes based on the ground state degeneracy (GSD), $\Tilde{G}_{\lambda}$, at the
extermum of the interpolation, with the interpolation strength given by $\lambda$:

\begin{enumerate}
    \item $\Tilde{G}_{\lambda=0} \neq \Tilde{G}_{\lambda=1}$ 
    \item $\Tilde{G}_{\lambda=0}$ = $\Tilde{G}_{\lambda=1}$ 
\end{enumerate}

\section{TPT'\lowercase{s}: \boldmath{$\Tilde{G}_{\lambda=0} \neq \Tilde{G}_{\lambda=1}$} \label{sec_ugsd}}
The phase transitions in this section are characterized by the change in the GSD of the frustration free
Hamiltonians at either end of the interpolation. We present such phase transitions induced by, both, change in
topology and change in boundary conditions.

\subsection{Topology variation: Torus with no domain wall to a cylinder with a mixed boundary\label{tvugsd}}

\begin{figure}[h!]
\includegraphics[width=\linewidth]{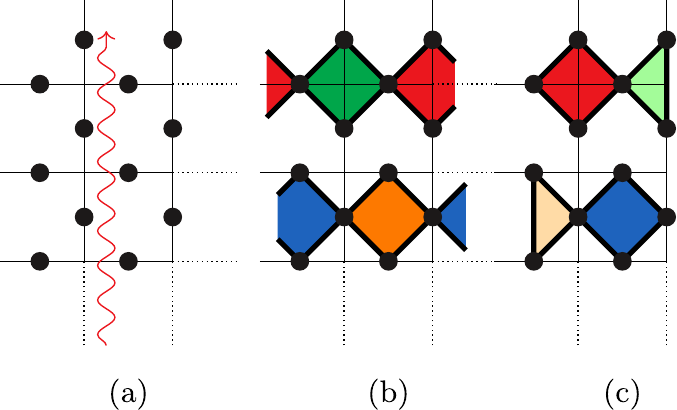}
\caption{\label{fig1}(a) The red snake represents the interpolation cut. (b) TC with periodic
boundaries i.e., on a torus. (c) TC with mixed boundaries on a cylinder. The red (blue) diamond represents
the $A_{v}^{\mathbin{\blacklozenge}}$ ($B_{p}^{\mathbin{\blacklozenge}}$) interaction whose interaction 
strength is unperturbed by the interpolation. As a result of interpolation the dark green (orange) full 
diamonds get mapped to light green (light orange) half diamonds and thereby the interaction is given by
$(1-\lambda)A_{v}^{\mathbin{\blacklozenge}} - \lambda A_{v}^{\Ltri}$,
$[(1-\lambda)B_{p}^{\mathbin{\blacklozenge}}-\lambda B_{p}^{\Rtri}]$.}
\end{figure}

By tuning the  local interactions, we map the TC Hamiltonian on a torus to a TC Hamiltonian on a cylinder
with mixed boundaries. The tuning breaks the periodicity of the torus and effectively gives rise to a 
cylinder with different open boundaries at either end, as in Fig.~\ref{fig1}(a). The interpolating Hamiltonian
connecting the different underlying topologies is given by Eq.~\ref{eq1}.

\be\label{eq1}
\begin{gathered}
\begin{split}
 H_{pm}(\lambda) & =  -\sum_{v}A_{v}^{\mathbin{\blacklozenge}}
            -\sum_{p}B_{p}^{\mathbin{\blacklozenge}} \\
	    & -(1-\lambda)\sum_{v'}A_{v'}^{\blacklozenge}
	    -(1-\lambda)\sum_{p'}B_{p'}^{\blacklozenge} \\
	    & -\lambda\sum_{v'}A_{v'}^{\Ltri}
	    -\lambda\sum_{p'}B_{p'}^{\Rtri},
\end{split}
\end{gathered}
\ee
where $A_{v}^{\mathbin{\blacklozenge}} = \prod\limits_{i=1}^{4}\sigma_{x}^{(i)}$ 
($B_{p}^{\mathbin{\blacklozenge}} = \prod\limits_{j=1}^{4}\sigma_{z}^{(j)}$) act on the four edges 
attached to the respective vertices (faces) in the bulk, while $A_{v}^{\Ltri} = 
\prod\limits_{i=1}^{3}\sigma_{x}^{(i)}$ ($B_{p}^{\Rtri} = \prod\limits_{j=1}^{3}
\sigma_{z}^{(j)}$) act on the three edges attached to the respective vertices (faces) at the boundary, as 
elucidated in Fig.~\ref{fig1}(b), (c). 

From Eq.~\ref{eq1}, we infer that at $\lambda$ = 0, $H_{pm}(0)$, represents the TC Hamiltonian on torus while at 
$\lambda= 1$, $H_{pm}(1)$, represents the TC Hamiltonian on cylinder with mixed boundary conditions. As the system is
perturbed by varying $\lambda$ from 0 to 1, the GSD changes from 4 to 1, indicating the presence of a TPT. The above TPT
is accompanied by break in both parity conservation and anyonic symmetry, as in the limit of $\lambda=0$ both are conserved
while in the limit of $\lambda=1$ both the symmetries remain broken. We study the energy gap opening in the degenerate manifold,
Topological Entanglement Entropy (TEE) with  respect to different cuts and the expectation value of open-loop operator to
gain further insight into the nature of phase transition.

\subsubsection{Energy gap}
\begin{figure}
\includegraphics[width=\linewidth]{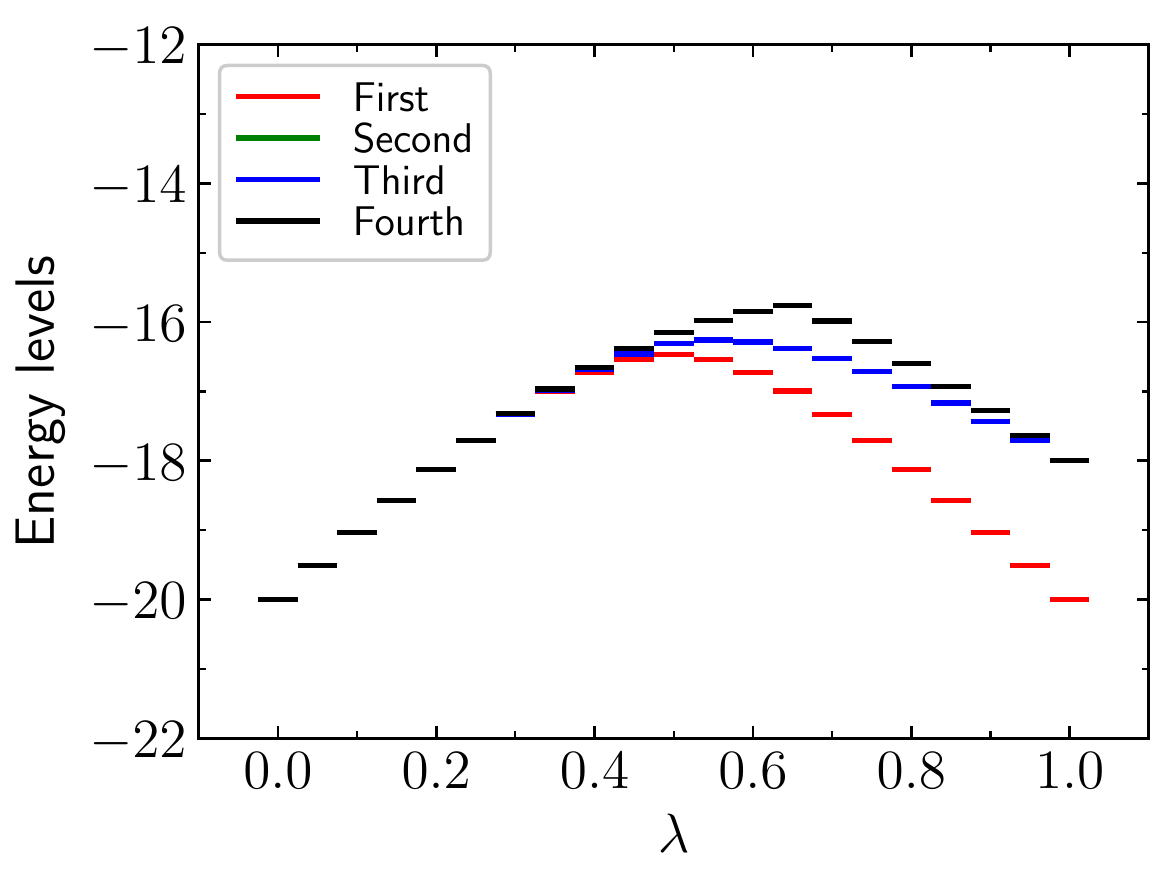}
\caption{\label{fig2}The least energy levels for a system size of $N=20$. At $\lambda=0$, we see that the ground state spectrum
is degenerate, while in the limit of $\lambda=1$ we have a unique ground state.}
\end{figure}

The ground state of the Hamiltonian, $H_{pm}(\lambda)$, both at $\lambda=0$ and at $\lambda=1$ is given by
$\mathcal{N}\prod\limits_v(\mathds{1} + A_{v})\ket{\textbf{0}}$, where the product is modified to include the vertices in
respective topologies and $\mathcal{N}$ is the normalization constant. In the limit of $\lambda=0$, the action of the
non-trivial loop operators around the legs of the torus maps between different degenerate ground states.
Since we consider a torus of genus one, the number of non-trivial loop operators are four, thereby the
GSD is 4. 
While in the  limit of $\lambda=1$, the non-trivial loop operator, along the periodic boundary of the cylinder,
leaves the ground state invariant, thereby we have a unique ground state \cite{Wang2015}. Therefore, for
some critical strength, $\lambda_{c}$, we expect a gap opening in the degenerate ground state spectrum, as in,
Fig.~\ref{fig2}. 

\begin{figure}
\includegraphics[width=\linewidth]{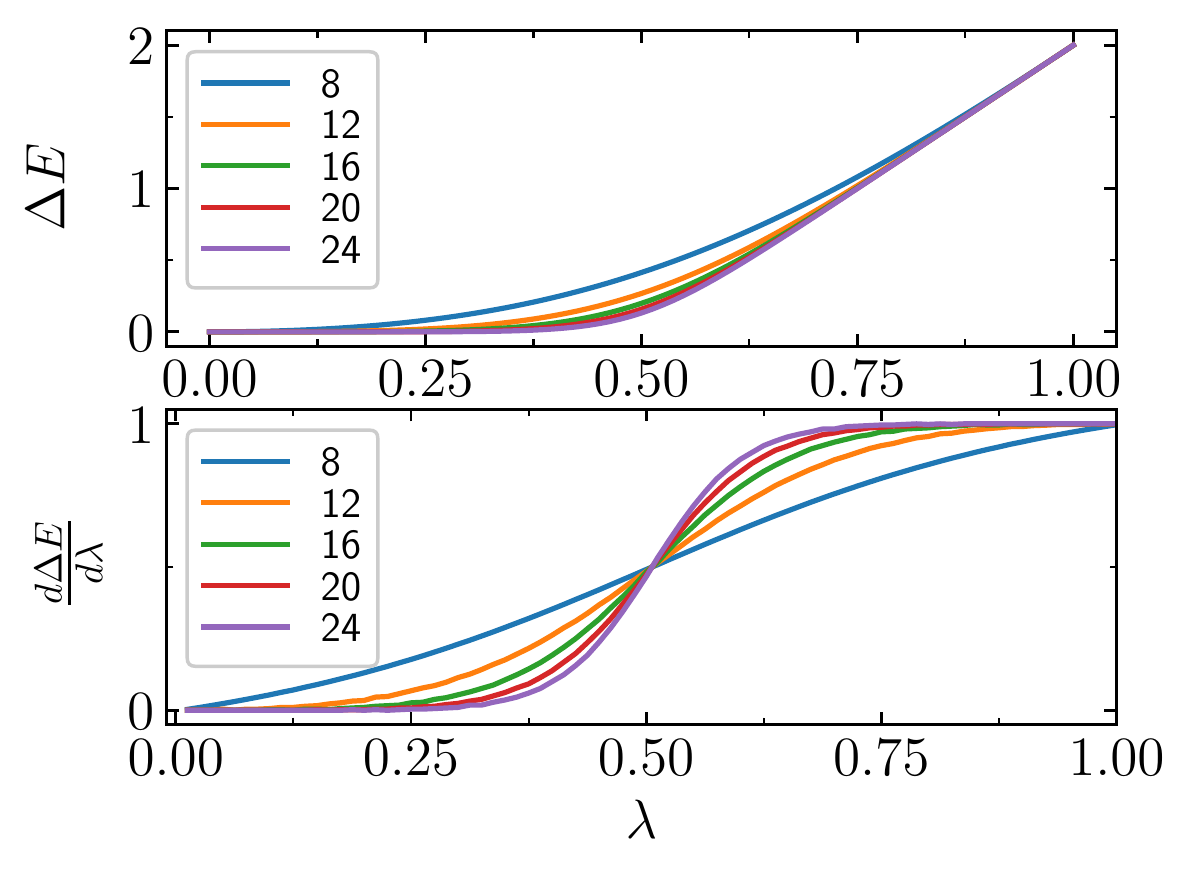}
\caption{\label{fig3}(Top) Difference between the least two energy levels, $\Delta E$ (Bottom) $\frac{d\Delta E}{d\lambda}$,
with the labels correponding to different system sizes.}
\end{figure}

From Fig.~\ref{fig3}, we note that there is a suppression in the energy gap $\Delta E$,
with increase in the system size, implying the ground state manifold is degenerate upto  a critical strength
and from its derivative we infer that the critical strength is around 0.5. We note that for the computation of relevant
low energy spectrum and relevant ground state properties we have used the linear algebra routines of Julia\cite{Bezanson2017}.

\subsubsection{Topological Entanglement Entropy}
\begin{figure}[h!]
\includegraphics[width=\linewidth]{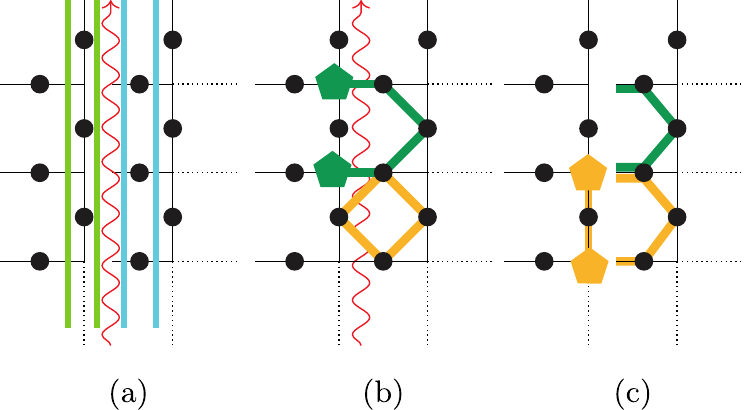}
\caption{\label{fig4}(a) The cuts used in the computation of TEE, the green and blue regions capture a strip on the
torus while in the  mixed boundary scenario, the green region captures the smooth boundary and the blue region
captures the rough boundary. (b) TC on a torus, the green string represents the $\sigma_{z}$ open-loop operator while the
golden string represents the trivial Wilson loop operator. (c) Due to the condensation of the excitation at the
boundary the green string reduces to a trivial open string while the Wilson loop splits into two open strings, one 
identical to the green string while the other sporting two excitations at its ends (excitations
are denoted by pentagons).}. 
\end{figure}

A key signature of topological order is the constant subleading term in the entanglement entropy, called
the Topological Entanglement Entropy (TEE), $\gamma$ \cite{Kitaev2006, Levin2006}. To compute $\gamma$, we refer to the procedure
outlined in Ref.~\cite{Jamadagni2018}. The cut used in the computation of $\gamma$ scales with the radius of the mixed 
boundary cylinder as in Fig.~\ref{fig4}(a). 

\begin{figure}
\centering
\includegraphics[width=\linewidth]{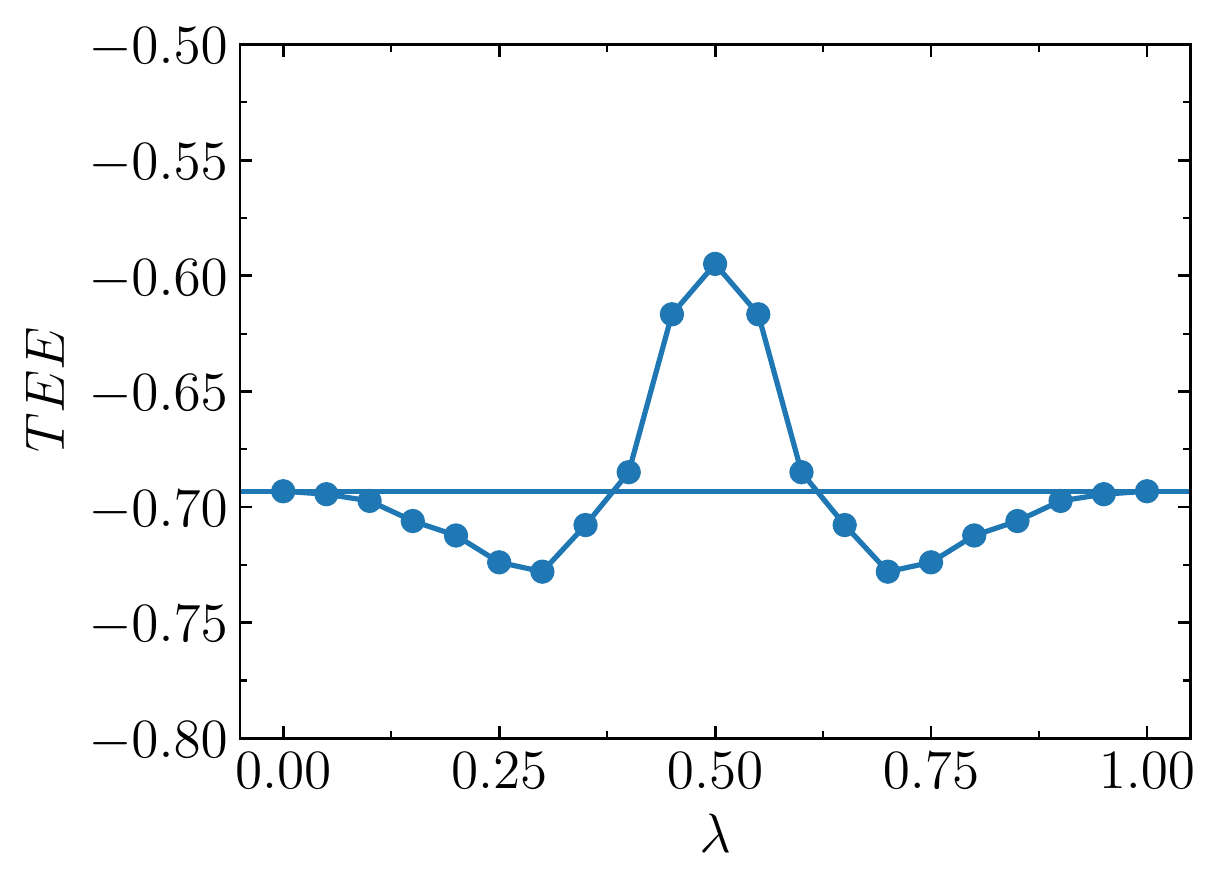}
\caption{\label{fig5}Topological Entanglement Entropy (TEE) as a function of the interpolation strength,
$\lambda$.}
\end{figure}

From Fig.~\ref{fig5}, we note that the TEE is around $\log2$ for 
all $\lambda$ and attribute the deviation from $\log2$ to finite size effects, as reported earlier in 
Ref.~\cite{Morampudi2014}. We further strengthen the claim from the above reference, that TEE is ineffective 
in detecting a phase transition between two different topological phases.

\subsubsection{Open-loop operator}
\begin{figure}
\centering
\includegraphics[width=\linewidth]{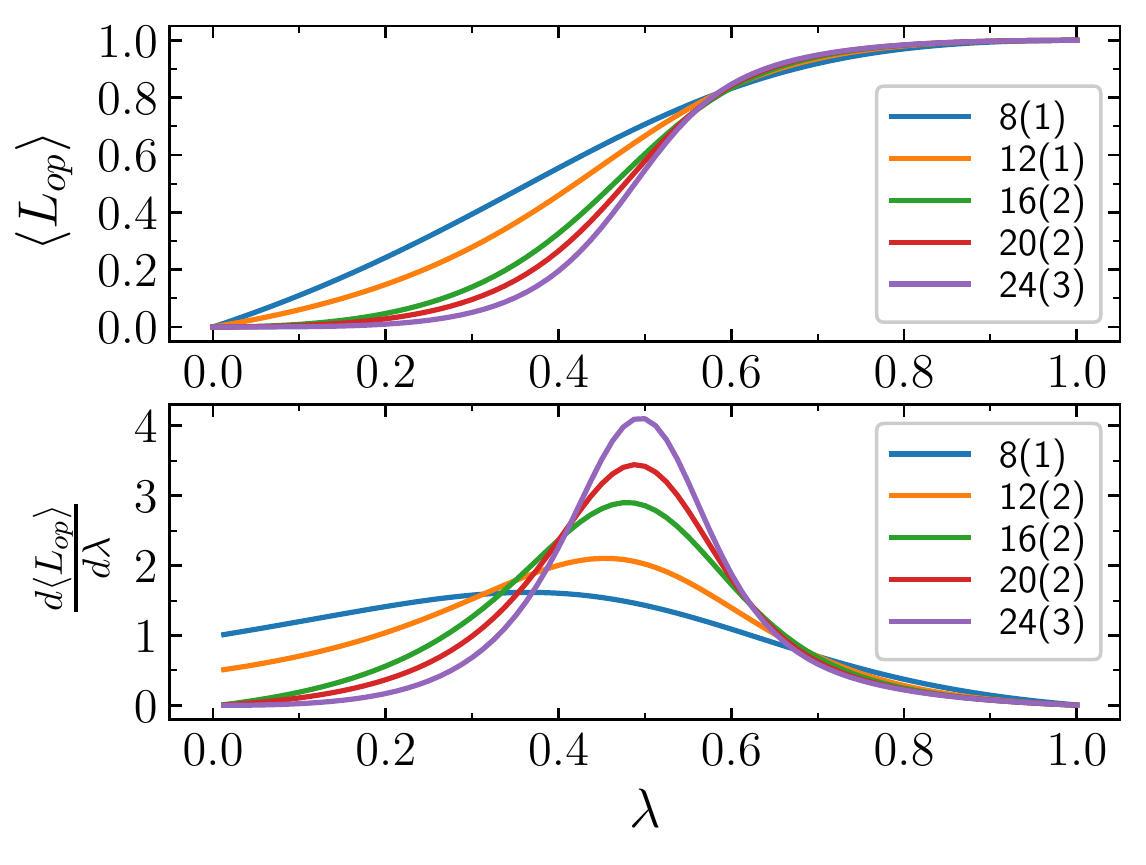}
\caption{\label{fig6}(Top) Expectation value of the longest open-loop operator (Bottom) Derivative of the 
expectation value with respect to $\lambda$. The labels denote different system sizes with the value in the 
parentheses indicating the maximal possible separation between the excitations used for the construction of the  
longest open loop operator.}
\end{figure}

We introduce the open-loop operator as in Fig.~\ref{fig4}(b) with periodic boundary as the
reference. The open-loop operators are generated by a sequence of $\sigma_{z}^{(i)} (\sigma_{x}^{(j)})$ operators
and are marked with excitations at their ends. Let us consider the open-loop operator as in Fig.~\ref{fig4}(b),
the expectation value with respect to the ground state at $\lambda=0$ is zero, i.e.,
$\braket{\psi_{gs}^{\lambda=0}|L_{z}^{r}|\psi_{gs}^{\lambda=0}}$ = 0, as the loop operator
projects the ground state into an excited state. While on the other hand at $\lambda=1$,
$\braket{\psi_{gs}^{\lambda=1}|L_{z}^{r}|\psi_{gs}^{\lambda=1}}$ = 1, since the 
excitations at the end of the open-loop condense on the boundary leaving the ground state invariant. 
We note that the expectation value of the longest open-loop operator i.e., the operator connecting
excitations which are maximally separated, effectively captures the phase transition.
From Fig.~\ref{fig6} and by performing finite size analysis, we infer that the expectation value
diverges at critical strength of $\lambda_{c}=0.533 \pm .032$, thereby signalling a phase transition.

\subsection{Boundary variation: Cylinder with rough boundaries to a mixed boundary \label{tcbv}}

\begin{figure}[h!]
\includegraphics[width=0.75\linewidth]{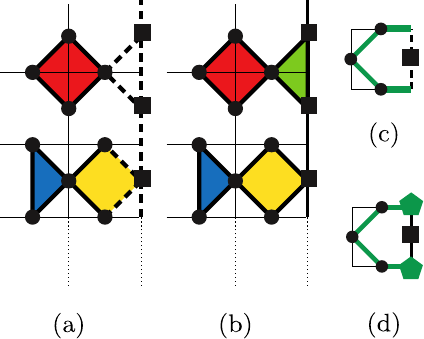}
\caption{\label{fig7}(a) TC on a cylinder with a rough boundary on both ends ($\lambda=0$).  Additional spins are
added on the right boundary, represented by $\blacksquare$.  (b) TC on a cylinder with mixed boundaries.
In both (a), (b) the red diamond remains unperturbed with action on the attached 
edges given by $A_{v}^{\blacklozenge}$, the dark blue half diamond also remains unperturbed 
with the action on the attached edges given by $B_{p}^{\Rtri}$. The yellow diamond in (a) 
represents the $B_{p}^{\boxdot}$ which translates to $B_{p}^{\blacklozenge}$ in (b), while the
uncolored dashed half diamond in (a) maps to  $A_{v}^{\Ltri}$ in (b) due to the interpolation.
The action of open-loop operator at the  boundary at (c) $\lambda=0$, (d) $\lambda=1$.} 
\end{figure}

In this section, we consider the TC Hamiltonian on a cylinder and interpolate between rough boundary
on both ends to a mixed boundary. The phase transition is similar to topology interpolation case as the GSD varies
from 2 to 1 as we vary the interpolation strength. The phase transition is marked by the break in the parity
conservation of the $m$-type excitations, as at $\lambda=0$ the $m$-type excitations always appear in pairs while at
$\lambda=1$ single excitations can be drawn from the boundary. We also note that there is no anyonic symmetry present
in the limits of $\lambda=0$ and $\lambda=1$. We interpolate the right rough boundary to a smooth boundary while the left
boundary remains unperturbed, see Fig.~\ref{fig7}. To this extent, we decorate the right
boundary, $R$, with additional spins denoted by $\blacksquare$ as in Fig.~\ref{fig7} and thereby add additional 
terms to the Hamiltonian, like $B_{p}^{\boxdot}$, the projector ${\ket{0}\bra{0}}$  as in Eq.~\ref{eq2}, 
which facilitate the interpolation while effectively retaining the boundary properties.

\be\label{eq2}
\begin{gathered}
\begin{split}
 H_{rm}(\lambda) & =  -\sum_{v}A_{v}^{\mathbin{\blacklozenge}}
            -\sum_{p}B_{p}^{\Rtri} \\
	    &-(1-\lambda)\sum_{p \in R}B_{p}^{\boxdot}
	    -(1-\lambda)\sum_{\blacksquare \in R}\ket{0}\bra{0}\\
	    &-\lambda\sum_{v \in R}A_{v}^{\Ltri}
	    -\lambda\sum_{p \in R}B_{p}^{\mathbin{\blacklozenge}},
\end{split}
\end{gathered}
\ee

where $A_{v}^{\mathbin{\blacklozenge}}$, $B_{p}^{\mathbin{\blacklozenge}}$, $A_{v}^{\Ltri}$,
$B_{p}^{\Rtri}$ are as defined in Sec.~\ref{tvugsd}. 
 . 
At $\lambda=0$, the above Hamiltonian reduces to the case of rough boundary at both open ends as the right
boundary spins are projected to $\ket{0}$ \cite{Beigi2011}, captured by the projector $\ket{0}\bra{0}$
and the typical $B_{p} = \prod\limits_{j}\sigma_{z}^{(j)}$ face interaction at the boundary has to be modified to
include the projection at the boundary and therefore modifies itself as $B_{p}^{\boxdot}$, given by
\be\label{eq3}
  B_{p}^{\boxdot} = \frac{1}{2}(\mathds{I}^{\bullet}\mathds{I}^{\bullet}\mathds{I}^{\bullet} + 
  \sigma_{z}^{\bullet}\sigma_{z}^{\bullet}\sigma_{z}^{\bullet})
  (\frac{\mathds{1} + \sigma_{z}}{2})^{\blacksquare}
\ee
where $\bullet$ indicates the action on the spins from the bulk and $\blacksquare$ indicates the action on 
the spin of the boundary.

\subsubsection{Energy gap}
\begin{figure}
	\includegraphics[width=\linewidth]{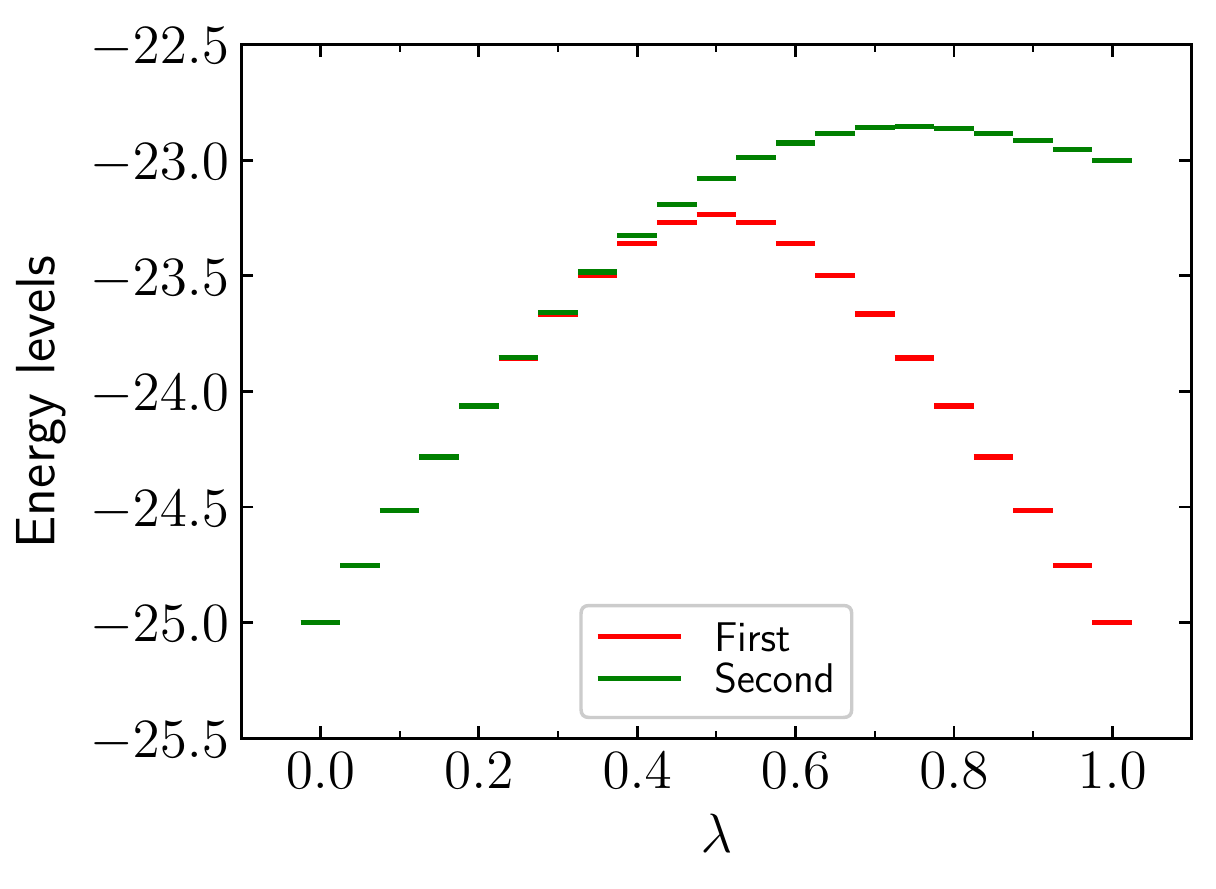}
	\caption{\label{fig8}Low energy spectrum of the interpolating Hamiltonian $H_{rm}(\lambda)$ for a system size of 
	$N=20$ spins.}
\end{figure}

At $\lambda=0$ and at $\lambda=1$, using the fact that the ground state is a simultaneous
ground state of all the operators in the Hamiltonian, one of the ground state can be represented as
$\mathcal{N}\prod\limits_v(\mathds{1} + A_{v})\ket{\textbf{0}}$, with the product modified suitably to include
vertices depending on the value of $\lambda$. In the limit of $\lambda=0$, the ground state manifold is double
degenerate \cite{Wang2015}, 
while in the limit of $\lambda=1$, the ground state is unique, see Fig.~\ref{fig8}.
In addition we note that the nature of the energy difference plot, $\Delta E$ versus $\lambda$, is similar to
Fig.~\ref{fig3} with the critical strength around 0.5.

\subsubsection{Open-loop operator}
\begin{figure}
	\centering
	\includegraphics[width=\linewidth]{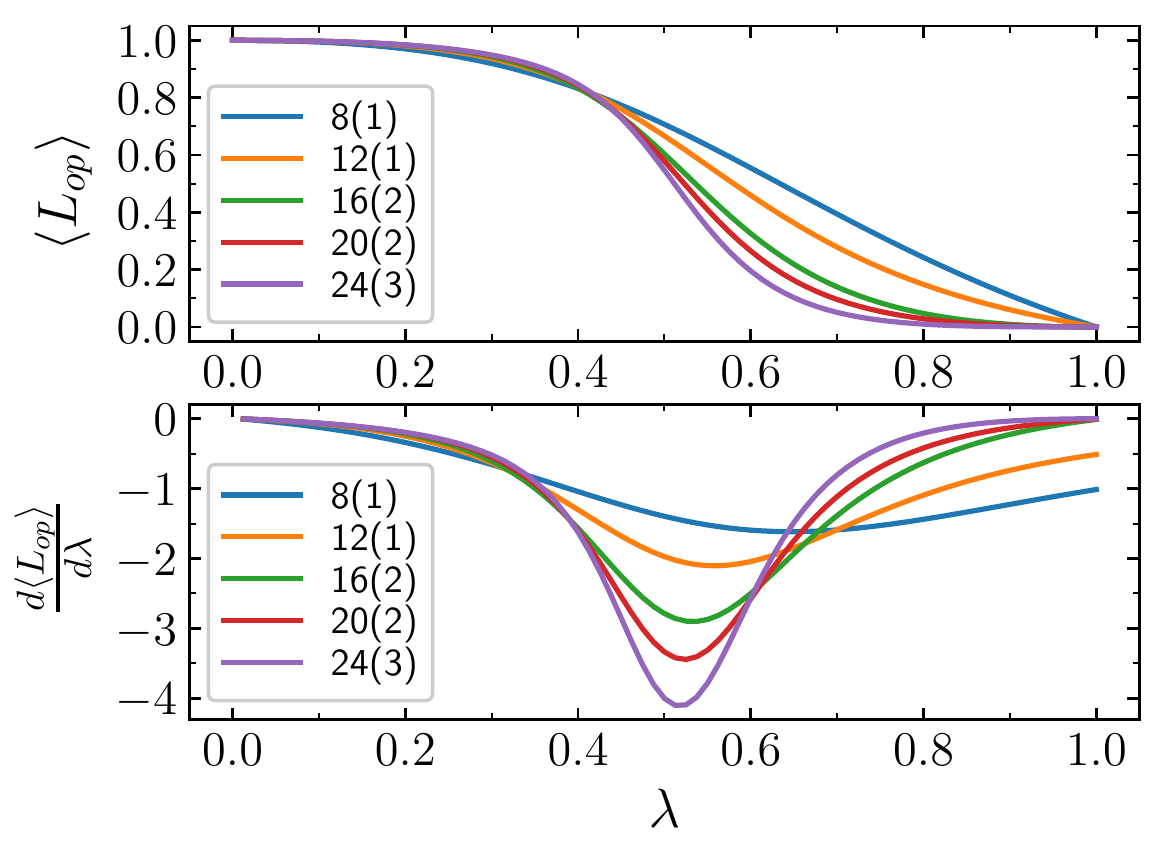}
	\caption{\label{fig9}(Top) Expectation value of the longest open-loop operator with respect to the interpolation 
	strength, $\lambda$. (Bottom) Derivative of the expectation value of the open-loop operator with respect to
	$\lambda$. The labels denote the different system sizes with the value in the parantheses as defined above in
	Fig.~\ref{fig6}.}
\end{figure}

As in the topology variation case, we compute the expectation value of the longest open-loop operator.
With reference to the rough boundary, the open-loop operator has excitations condensing at the boundary at 
$\lambda=0$, therefore the expectation value is 1, where as at $\lambda=1$ the excitations are retained at the
boundary, see Fig.~\ref{fig7}(c), (d), with the expectation value going to zero. From Fig.~\ref{fig9} and by performing
finite size analysis we note that the expectation value diverges at $\lambda_{c}=0.481 \pm 0.048$.

\section{TPT'\lowercase{s}: \boldmath{$\Tilde{G}_{\lambda=0} = \Tilde{G}_{\lambda=1}$} \label{sec_egsd}}
In this section, we introduce various scenarios where the phase transitions are characterized by closing of the
energy gap between the ground state manifold and the first excited state along the path of interpolation. We
investigate for such cases in the context of topology variation as well as boundary variation.

\subsection{Topology variation: Torus with domain wall to a cylinder with rough boundaries}
We briefly motivate the notion of domain wall as one of the boundaries of the TC and then further discuss the
presence of TPT as we dissect the torus along the domain wall to a cylinder with rough boundaries at either end.

\begin{figure}[h!]
\includegraphics[width=\linewidth]{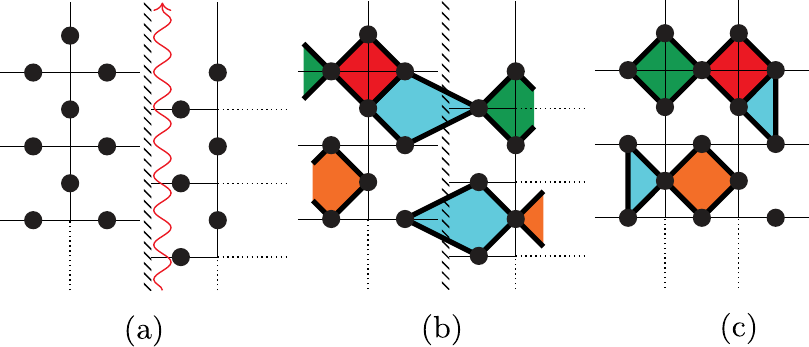}

\vspace{0.5cm}

\includegraphics[width=0.9\linewidth]{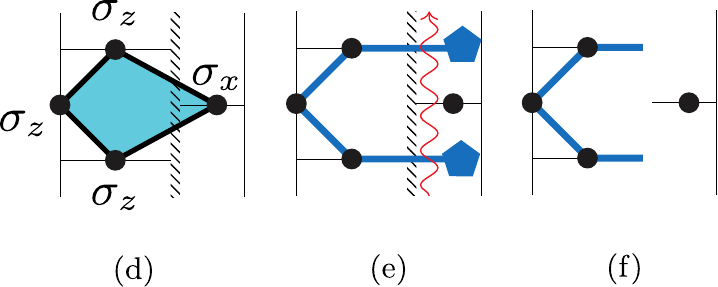}
\caption{\label{fig10}(a) The interpolation cut, denoted by the red snake dissects along the domain wall. (b) At
$\lambda=0$, TC on a torus with a domain wall, denoted by the short slant interface. (c) At $\lambda=1$, TC on a
cylinder with a rough boundary on both ends. (d) $B_{p}^{\boxplus}$ operator at the domain wall. (e) Open-loop
operator with a pair of excitations projecting the ground state at $\lambda=0$ into an excited state. (f) Open-loop
operator whose excitations have condensed at the boundary leaving the ground state at $\lambda=1$ invariant under
the loop action.}
\end{figure}

The authors in Ref.~\cite{Beigi2011} have introduced the notion of domain walls between two different TO
phases, given by the quantum doubles $D(G_{1})$, $D(G_{2})$. Further, it has been shown that the domain walls between
such quantum doubles are equivalent to the boundary conditions of the folded quantum double $D(G_{1}\times G_{2})$,
which are characterized by the subgroups, $K$, of $G_{1}\times G_{2}$, along with a non-trivial 2-cocycle of $K$. In
the case of folded toric code which is given by $D(Z_{2}\times Z_{2})$, there exists a domain wall given by the
subgroup $Z_{2}\times Z_{2}$ along with a non-trivial 2-cocycle of $Z_{2}\times Z_{2}$ which when unfolded reduces to
a boundary as illustrated in Fig.~\ref{fig10}(b). The Hamiltonian of the TC with a domain wall is given by $H_{dr}(0)$, as in
Eq.~\ref{eq4}. The modified $B_{p}$ operator at the domain wall, $B_{p}^{\boxplus}$, takes the form as in
Fig.~\ref{fig10}(d)\cite{Kitaev2012, Yoshida2017}. The interpolating Hamiltonian connecting the TC with a domain wall on torus
to TC on a cylinder with rough boundaries
is given by Eq.~\ref{eq4}
\be\label{eq4}
\begin{gathered}
\begin{split}
 H_{dr}(\lambda) & =  -\sum_{v}A_{v}^{\mathbin{\blacklozenge}}
            -\sum_{p}B_{p}^{\mathbin{\blacklozenge}}\\
            & -(1-\lambda)\sum_{p'}B_{p}^{\boxplus}\\
	    &-\lambda\sum_{p''}B_{p}^{\Ltri}
	    -\lambda\sum_{p''}B_{p}^{\Rtri},
\end{split}
\end{gathered}
\ee
where $A_{v}^{\mathbin{\blacklozenge}}$, $B_{p}^{\mathbin{\blacklozenge}}$, $B_{p}^{\Rtri}$ are
defined as in Sec.~\ref{tvugsd}, while $B_{p}^{\Ltri}$ is qualitatively identical to
$B_{p}^{\Rtri}$. The phase transition is characterized by break in the parity and anyonic symmetry. The
parity of the $m$-type excitations is preserved in the limit of $\lambda=1$ while is broken in the limit of $\lambda=0$.
On the other hand, anyonic symmetry is preserved in the limit of $\lambda=0$ and is broken in the limit $\lambda=1$.

\subsubsection{Energy gap}
\begin{figure}[h!]
\includegraphics[width=\linewidth]{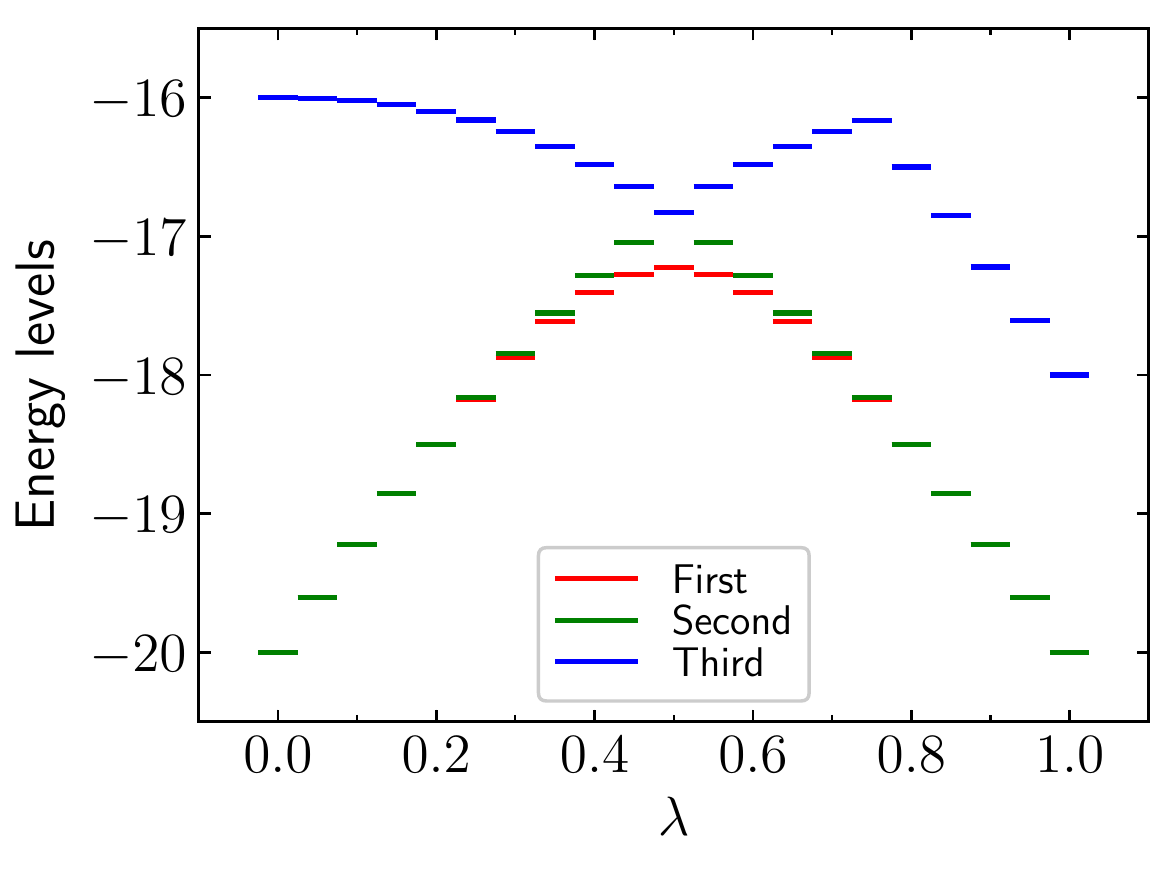}
\caption{\label{fig11}The least energy levels for a system size of $N=20$. At $\lambda=0$ and $\lambda=1$, we note that the 
ground state manifold is degenerate, while around $\lambda=0.5$, we note the split in the degeneracy along with the
the merging of the first and second excited states.}
\end{figure}

At both $\lambda=0$ and $\lambda=1$, the ground state manifold is two fold degenerate. Using the notion established in
the earlier sections, one of the representations of the ground state at $\lambda=0$ is given by
$\mathcal{N}\prod_{v}(\mathds{1} + A_{v})\prod_{p}(\mathds{1} + B_{p}^{\boxplus})\ket{\textbf{0}}$, where as at
$\lambda=1$,  is given by $\mathcal{N}\prod_{v}(\mathds{1} + A_{v})\ket{\textbf{0}}$. In the limit of $\lambda=0$, the
other ground state can be obtained by the action of the non-trivial loop operator running parallel to the
domain wall. The other non-trivial loop operator running perpendicular to the domain wall does not leave the ground
state invariant as $m$-type violations get identified as $e$-type violations as they pass through the domain wall, the
fusion of which results in a fermion, instead of vacuum, as in the absence of the domain wall. Therefore, establishing
the fact that the GSD of the TC with a domain wall on torus is two. 
From Fig.~\ref{fig11}, for finite size system of $N=20$ spins, we see a split in the ground state manifold around
$\lambda=0.5$ and also note that the first and the second excited states merge. 

\begin{figure}
\includegraphics[width=\linewidth]{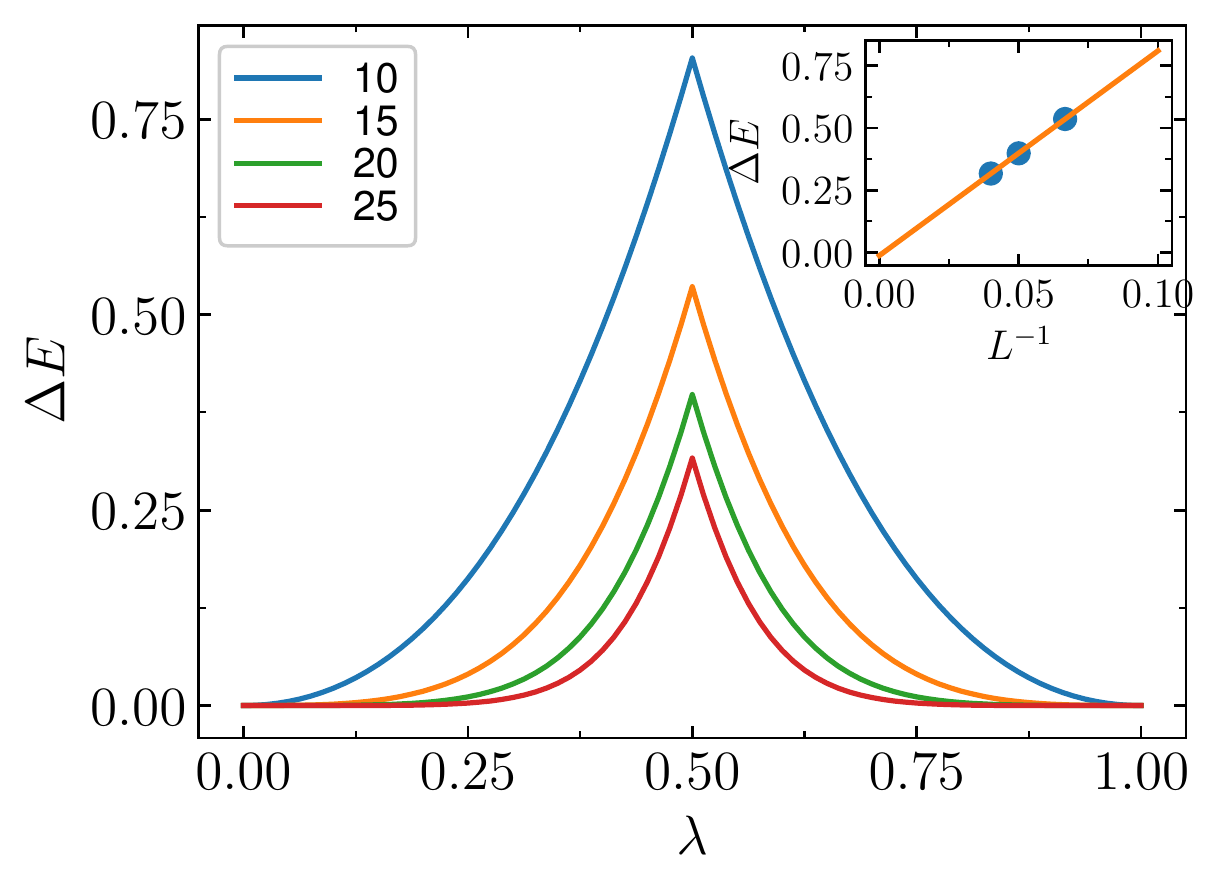}
\caption{\label{fig12}Energy difference between the first two energy levels as a function of the interpolation
strength, $\lambda$, with the labels denoting the different system sizes (Inset) Extrapolating the energy difference at 
$\lambda=0.5$, to the thermodynamic limit by performing finite-size analysis.} 
\end{figure}

\begin{figure}[h!]
\includegraphics[width=\linewidth]{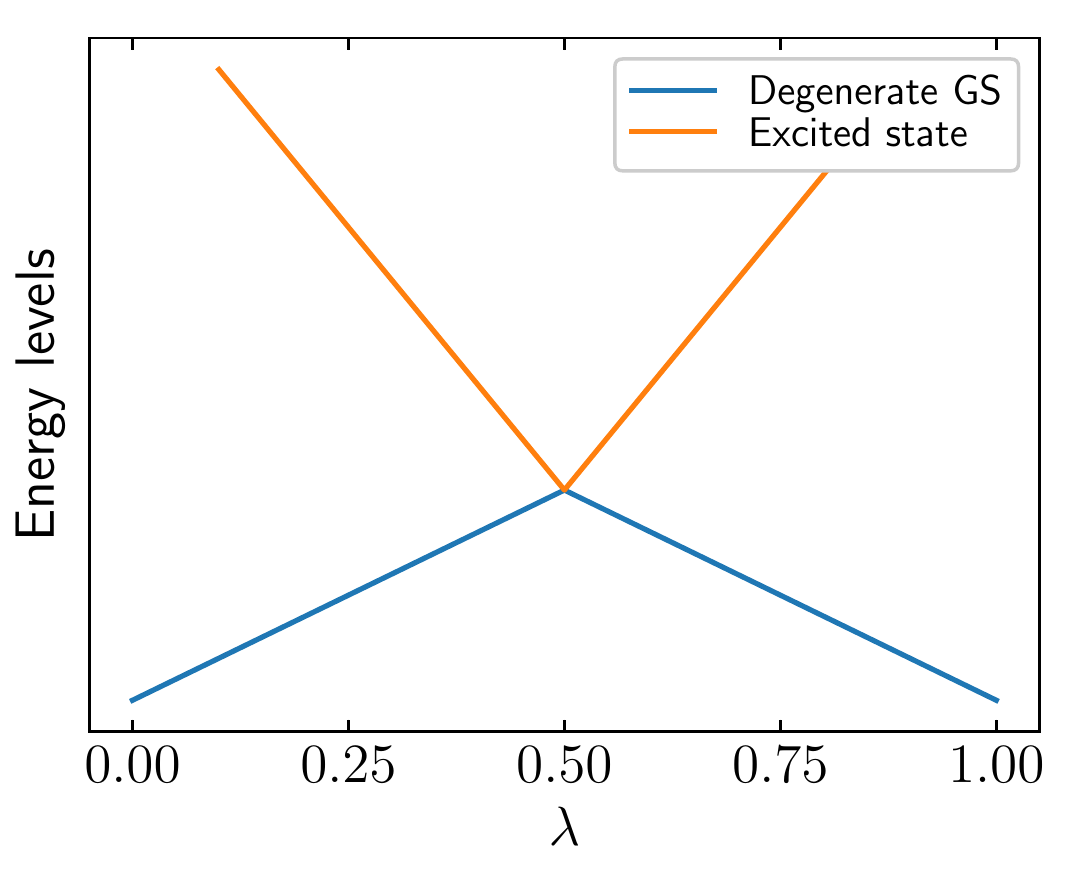}
\caption{\label{fig13}The potential energy spectrum in the thermodynamic limit as a function of the interpolation
strength, $\lambda$. The gap closing between the degenerate ground state manifold and the first excited state
indicates the presence of the phase transition.}
\end{figure}

From Fig.~\ref{fig12}, we note that the
energy gap between the ground state and the first excited state decreases with increase in system size. Extrapolating
to the thermodynamic limit by performing finite size analysis, we note that the degeneracy of the ground state manifold
is retained at all $\lambda$ and combining the fact that there is a energy gap closing at $\lambda=0.5$ results in a
energy spectrum as in Fig.~\ref{fig13} indicating the presence of a TPT at $\lambda=0.5$. 

\subsubsection{Open-loop operator}
\begin{figure}
\includegraphics[width=\linewidth]{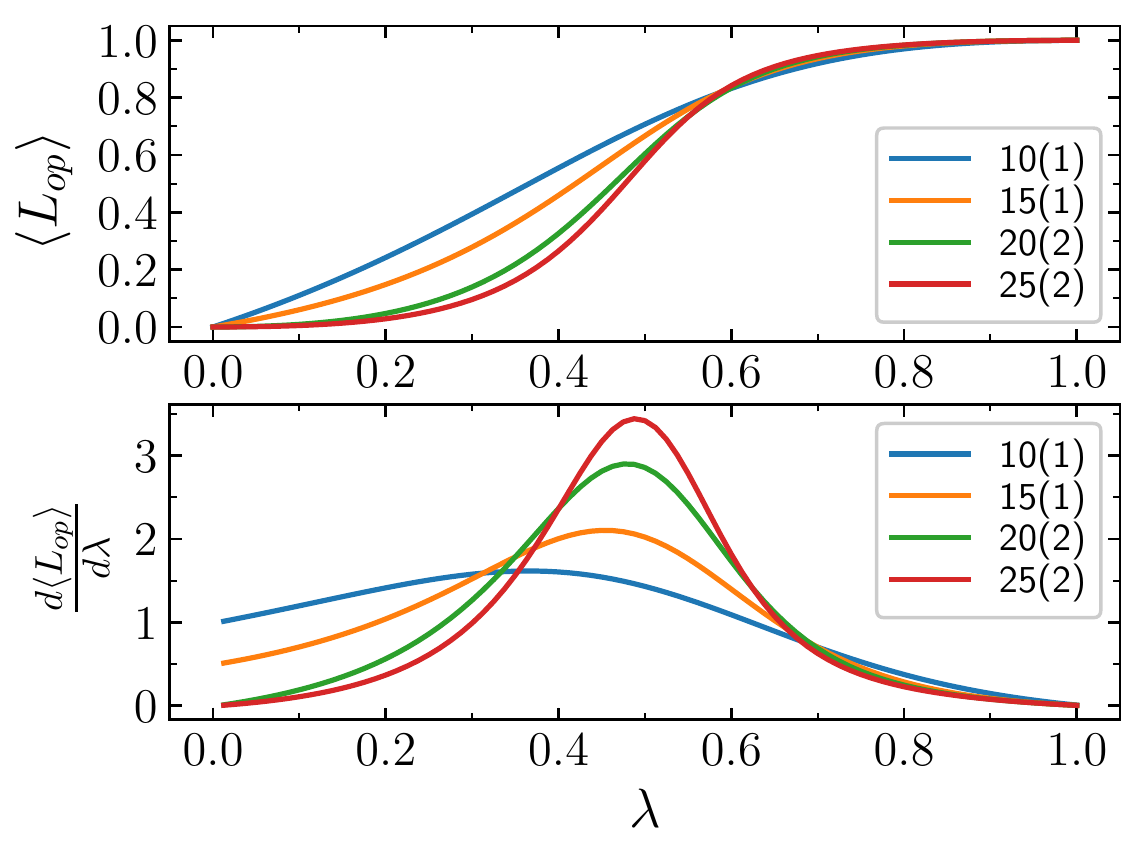}
\caption{\label{fig14}(Top) Expectation value of the longest open-loop operator with respect to different 
interpolation strength, $\lambda$. (Bottom) Derivative of the expectation value with respect to $\lambda$.
The labels denote the different system sizes with the value in the parantheses as defined earlier.}
\end{figure}

To further consolidate the presence of TPT, we compute the expectation value of the longest open-loop operator at
different interpolation strength, $\lambda$. We define the loop operator with reference to the TC on a torus with a
domain wall as in Fig.~\ref{fig10}(e). The open-loop is generated by the action of a sequence of $\sigma_{z}$
operators and sports two $B_{p}$ violations at its end. In this limit of $\lambda=0$, the loop operator projects the
ground state into an excited state, thereby leading to an expectation value of zero. While at the other extreme,
$\lambda=1$, the excitations at the end of the open-loop condense at the boundary, as in Fig.~\ref{fig10}(f), thereby 
leaving the ground state invariant and hence the expectation value is one in the vicinity of $\lambda=1$. From
Fig.~\ref{fig14} and by performing finite size scaling analysis we conclude that the critical strength
is given by $\lambda_{c} = 0.539 \pm 0.046$,

\subsection{Boundary variation: Cylinder with rough boundaries to smooth boundaries}
\begin{figure}[h!]
\includegraphics[width=0.75\linewidth]{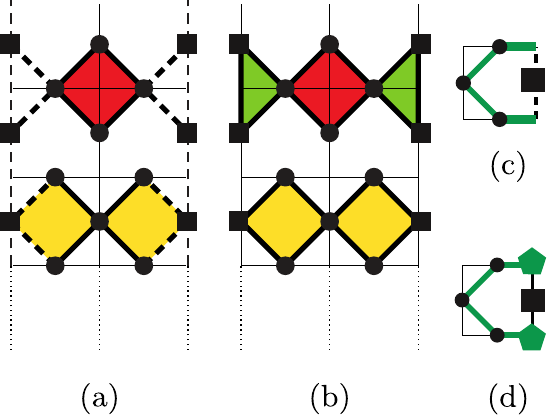}
\caption{\label{fig15}TC on a cylinder with (a) rough boundary, (b) smooth boundary on both ends. The red diamond
belongs to the interior region, $I$, which remains unperturbed while the transparent half
diamonds in (a) translate to half filled green diamonds $A_{v}^{\Ltri}$,
$A_{v}^{\Rtri}$ at either boundaries $L$ and $R$ respectively as $\lambda$ varies from 0 to 1.
Similarly, the golden yellow diamonds represent $B_{p}^{\boxdot}$ in (a) and map to
$B_{p}^{\mathbin{\blacklozenge}}$ in (b) with increase in $\lambda$. The action of the open-loop operator
at the boundary at (c) $\lambda=0$, (d) $\lambda=1$.} 
\end{figure}

In this section we present the boundary variation of the above TPT. To this extent, we interpolate between rough
boundary on both ends to smooth boundary on both ends of the cylinder, see Fig.~\ref{fig15}(a), (b). The interpolating
Hamiltonian is given by $H_{rs}$, as in Eq.~\ref{eq5}.

\be\label{eq5}
\begin{gathered}
\begin{split}
 H_{rs}(\lambda) &=  -\sum_{v \in I}A_{v}^{\mathbin{\blacklozenge}}\\
	    &-(1-\lambda)\sum_{p \in R}B_{p}^{\boxdot}
	    -(1-\lambda)\sum_{p \in L}B_{p}^{\boxdot}\\
	    &-(1-\lambda)\sum_{\blacksquare \in R}\ket{0}\bra{0}
	    -(1-\lambda)\sum_{\blacksquare \in L}\ket{0}\bra{0} \\
	    &-\lambda\sum_{v \in R}A_{v}^{\Ltri}
	    -\lambda\sum_{v \in L}A_{v}^{\Rtri}\\
	    &-\lambda\sum_{p \in R}B_{p}^{\mathbin{\blacklozenge}}
	    -\lambda\sum_{p \in L}B_{p}^{\mathbin{\blacklozenge}},
\end{split}
\end{gathered}
\ee
where $I$ denotes the interior bulk region, $R$ denotes the right boundary and $L$ denotes the left boundary. The phase
transition is characterized by break in the parity conservation of the  $m(e)$-type excitations. In the limit of
$\lambda=0$, $m(e)$-type excitations occur in pairs (singly) while in the limit of $\lambda=1$, $m(e)$-type excitations
appear singly (in pairs). There is no anyonic symmetry present in either phases due to the condensation at the boundary
i.e., the fusion rules are not invariant under the exchange of $e$ and $m$ labels.

\subsubsection{Energy gap}
\begin{figure}
	\includegraphics[width=\linewidth]{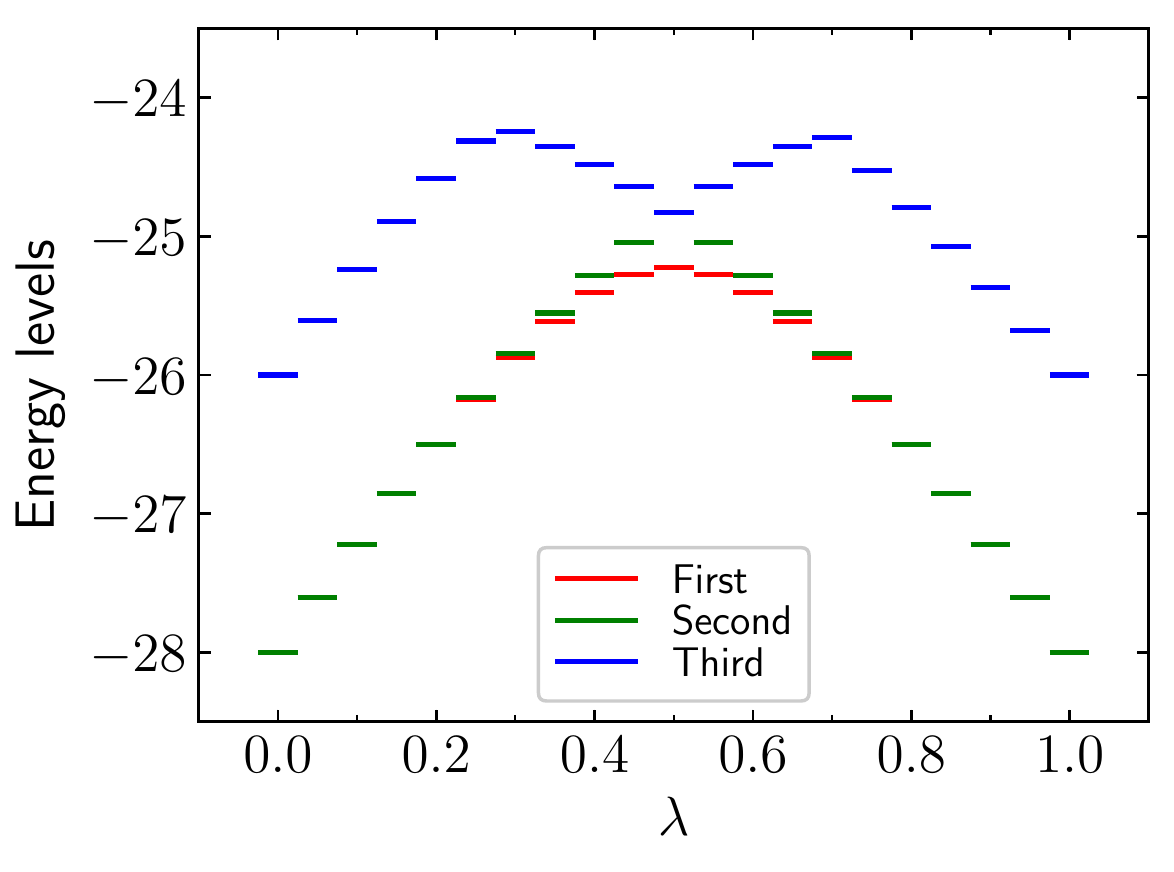}
	\caption{\label{fig16}Least energy levels for a system size of $N=20$ spins. Both at $\lambda=0$ and 
	$\lambda=1$, the ground state manifold is degenerate. At $\lambda=0.5$, we note the merging of the first and 
	the second excited energy levels.}
\end{figure}

The ground state manifold is two fold degenerate at the extremities of the interpolation parameter, $\lambda$ 
\cite{Wang2015}. As in the case of topology variation, it is evident that for finite size systems
the first and the second excited states merge at $\lambda=0.5$, see Fig.~\ref{fig16}. The energy difference 
between the first two energy levels is qualitatively similar to Fig.~\ref{fig12} and thereby in the thermodynamic 
limit the energy spectrum qualitatively resembles Fig.~\ref{fig13}, implying the presence of a phase transition
due to the closure of the energy gap. 

\subsubsection{Open-loop operator}
\begin{figure}
	\includegraphics[width=\linewidth]{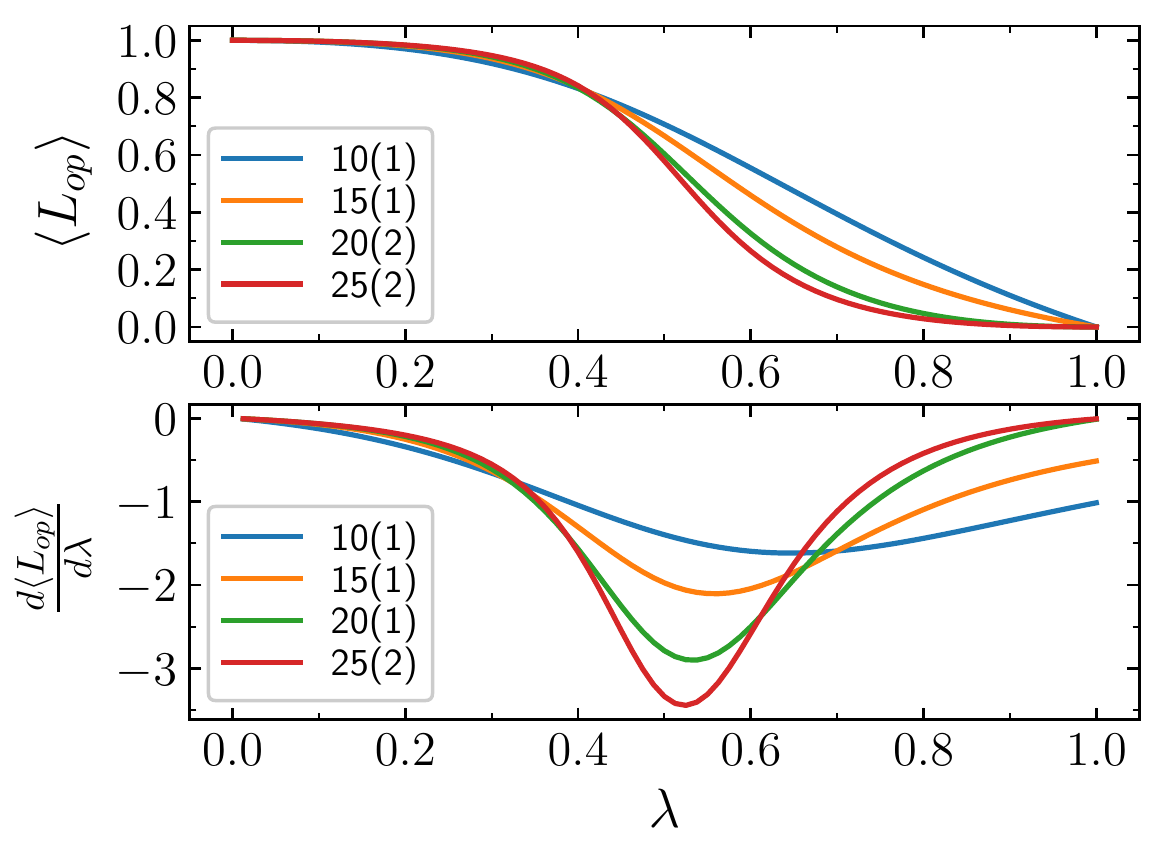}
	\caption{\label{fig17}(Top) Expectation value of the longest open-loop operator with respect to $\lambda$.
	(Bottom) Derivative of the expectation value with respect to $\lambda$. As noted earlier, the labels denote 
	the different system sizes.}
\end{figure}

Taking cue from the above analysis, we compute the expectation value of the open-loop operator to estimate the 
critical strength at which the phase transition occurs. The open-loop operator is generated by a sequence of 
$\sigma_{z}$ operators which holds $A_{v}$ excitations at its end. At $\lambda=0$, these excitations condense on the
boundary, while at $\lambda=1$, the excitations are retained at the boundary as in Fig.~\ref{fig15}(c), (d)
respectively. From Fig.~\ref{fig17}, and by performing finite size analysis we note that the expectation value 
diverges at $\lambda_{c}=0.463\pm 0.036$. 

\section{Interpolation via engineered dissipation \label{sec_idiss}}
We aim to achieve the interpolation introduced in Sec.~\ref{tvugsd}, in an open quantum system by engineering suitable 
collapse operators. To draw parallels with the closed system analysis, the study of phase transitions in open 
systems is associated with the properties of the steady states which are obtained by solving the 
Lindblad Master equation (LME)
\be
\begin{gathered}
\dot\rho(t) = -i[H(t),\rho(t)]  + \\
 \sum\limits_{n}\frac{1}{2}[2C_n\rho(t)C_n^\dag-\rho(t)C_n^\dag C_n-C_n^\dag C_n \rho(t)]
\end{gathered}
\ee
where $H$ is the Hamiltonian capturing coherent evolution while $C_{n}$'s are the collapse operators which encode the dissipative dynamics.

\begin{figure}[h!]
\includegraphics[width=\linewidth]{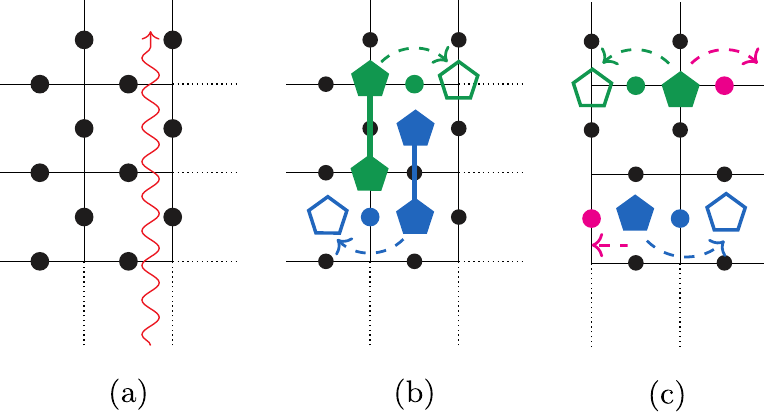}
\caption{\label{fig18}(a) The red snake represents the interpolation cut. The dissipative dynamics induced by the
collapse operators by diffusing excitations on (b) a torus (c) a cylinder with mixed boundaries. (b) 
Excitations always appear in pairs and the collapse operators diffuse the excitations (represented by dashed green and blue
arrows) or cool them by fusing (represented by thick green and blue lines). (c) Excitation parity is not conserved
because of the boundary, thereby allowing the excitations to condense at the boundary (represented by dashed magenta
arrows), in addition to the diffusion and pair cooling as noted in (b).}
\end{figure}

In Ref.~\cite{Weimer2010}, the authors have introduced collapse operators which cool a product state
to the entangled ground state of the TC. We consider a purely dissipative setup i.e., set $H$=0 and extend the above
construction, by introducing additional collapse operators whose effective cooling rate involves the interpolation
parameter, $\lambda$, thereby cooling to different ground states at the extremities of the interpolation. We analyze the case of
interpolation between the ground state of TC on a torus ($\lambda=0$) to the ground state on a cylinder with mixed
boundary conditions ($\lambda=1$) as introduced in Sec.~\ref{tvugsd}. For lucidity, we split the collapse operators into
three classes: the collapse operators acting on the permanent vertices (faces) given by $c_{v(f)}^{p}$, the collapse
operators acting on the periodic boundary given by $c_{v(f)}^{t}$ and the collapse operators acting on the open boundary
given by $c_{v(f)}^{o}$ and define them as in Eq.~\ref{cop}, Fig.~\ref{fig18}.

\begin{equation}\label{cop}
\begin{gathered}
\begin{split}
    c_{v}^{p} &= \frac{\sqrt{\gamma_{v}}}{2}\sigma_{z}^{(i)}(\mathds{1}-A_{v}^{\mathbin{\blacklozenge}}),\\
    c_{f}^{p} &= \frac{\sqrt{\gamma_{f}}}{2}\sigma_{x}^{(j)}(\mathds{1}-B_{f}^{\mathbin{\blacklozenge}}),\\
    c_{v}^{t}(\lambda) &= \frac{\sqrt{\gamma_{v}}}{2}(1-\lambda)\sigma_{z}^{(i)}(\mathds{1}-A_{v}^{\mathbin{
    \blacklozenge}}), \\
    c_{f}^{t}(\lambda) &= \frac{\sqrt{\gamma_{f}}}{2}(1-\lambda)\sigma_{x}^{(j)}(\mathds{1}-B_{f}^{\mathbin{
    \blacklozenge}}), \\
    c_{v}^{o}(\lambda) &= \frac{\sqrt{\gamma_{v}}}{2}\lambda\sigma_{z}^{(i)}(\mathds{1}-A_{v}^{\Rtri}), \\
    c_{f}^{o}(\lambda) &= \frac{\sqrt{\gamma_{f}}}{2}\lambda\sigma_{x}^{(j)}(\mathds{1}-B_{f}^{\Ltri}),
\end{split}
\end{gathered}
\end{equation}

where $\gamma_{v}, \gamma_{f}$ are the cooling rates of the vertex and face excitations, while $\lambda$ is the interpolation
strength, $A_{v}^{\mathbin{\blacklozenge}}$, $B_{f}^{\mathbin{\blacklozenge}}$, $A_{v}^{\Rtri}$,
$B_{f}^{\Ltri}$ operators are as defined in the earlier sections. Intuitively, the dynamics induced 
by the collapse operators diffuse the excitations around the lattice i.e., the excitations perform a random walk and upon
meeting another excitation or a relevant boundary, fuse, thereby cooling to a steady state. In the limit of $\lambda=0$ and
$\lambda=1$, the collapse operators effectively cool the product state to a pure steady state given by ground state of the
TC at respective $\lambda$. At intermediate $\lambda$, the dynamics is captured by the competition between the cooling
operators that promote the diffusion of the excitations along the periodic boundary and the cooling 
operators which promote a biased diffusion resulting in a restricted diffusion, effectively
capturing the break in topology. Due to the competitive cooling, the steady state at intermediate $\lambda$ is a mixed
state unlike the pure steady state at the extremities, hence the phase transition which we shall present shortly
is a mixed state phase transition. We further note that the phase transition analysis presented hereafter, is based on the
assumption that the steady state at all $\lambda$ is TO, thereby resulting in a TPT in an open system. 
The assumption can be substantiated by the fact that the mixed state obtained at intermediate $\lambda$, in the end, is 
due to a collective cooling scheme where the cooling itself is aimed at generating a TO pure state. We aim to present other
signatures for detecting QPT's between TO and trivial mixed states in a separate work and hence
the verification shall be postponed to the future\cite{Jamadagni2020}.

\begin{figure}
\includegraphics[width=\linewidth]{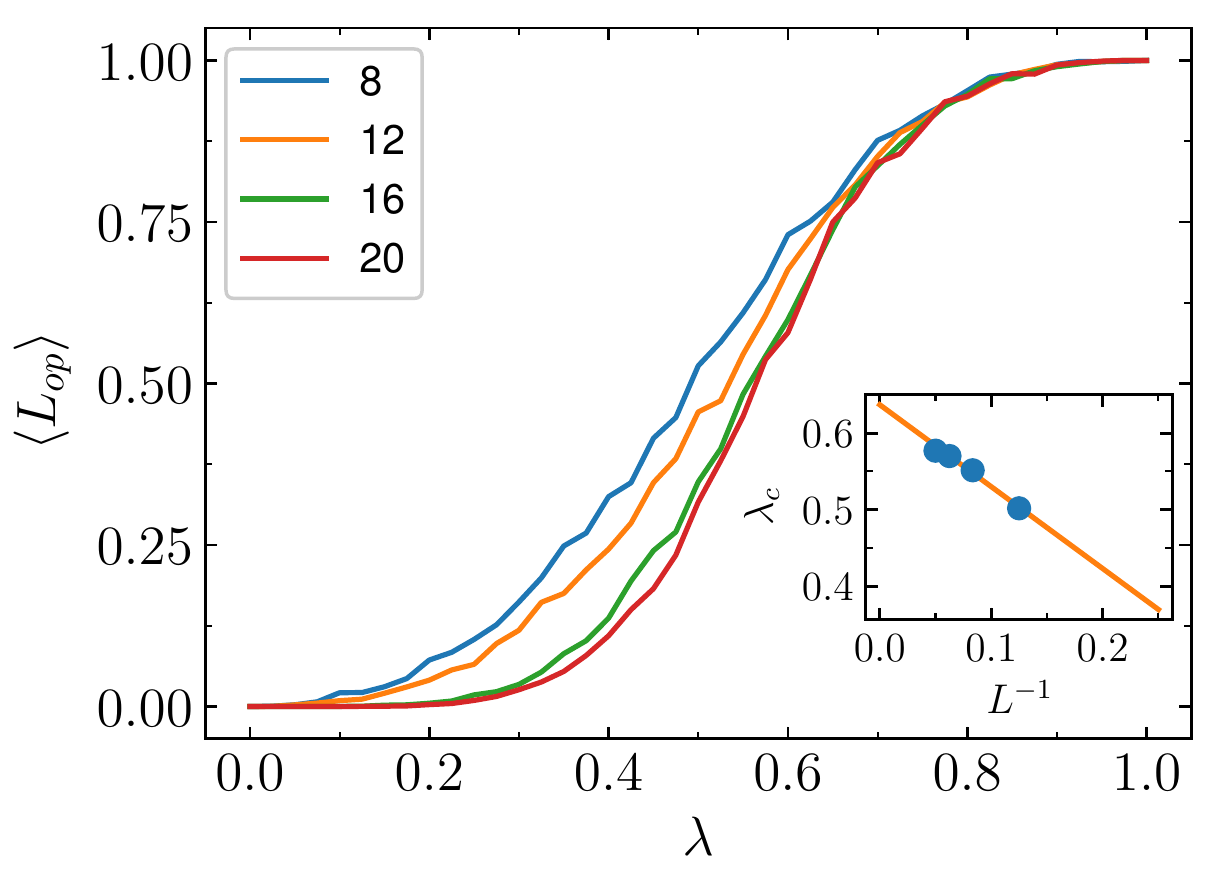}
\caption{\label{fig19}Expectation value of the longest open-loop operator with respect to the interpolation strength, 
$\lambda$ with the labels denoting the different system sizes. (Inset) Finite size scaling to obtain the critical strength,
$\lambda_{c}$.}
\end{figure}

We compute the steady states at different interpolating strength, $\lambda$, by using the Monte Carlo Wave 
Function (MCWF) method \cite{Johansson2013}. In the vicinity of $\lambda=0$, the dissipators cool the system to the
ground state of the TC on a torus while at $\lambda=1$, the dissipators cool the system to the ground state of the TC
on a cylinder. The expectation value of the open-loop operator, given by 
$\Tr({\rho_{\lambda} L})$ where $\rho_{\lambda}$ is the steady state at interpolation strength $\lambda$ and
$L$ is the open-loop operator, as in Fig.~\ref{fig4}(b), is used to distinguish the different topological phases. 
Using similar arguments presented earlier, we note that the expectation value of the open-loop operator 
is zero in the periodic boundary case where as is 1 in the open boundary case, with the critical strength at 
$\lambda_{c}=0.637 \pm.004$ obtained by performing finite size analysis, as in Fig.~\ref{fig19}.


\section{Summary and Discussion}
  In summary, we have studied the sensitivity of topological phases with respect to the boundary conditions of the underlying
manifold on which they are defined. We have considered the change in boundary conditions of two flavors: (a) effective
topology variation, where we have varied the underlying topology from periodic boundary to open boundary i.e., from torus
to a cylinder  (b) effective boundary variation, where we have fixed the underlying topology to a cylinder and have varied
the open boundaries of the cylinder. The sensitivity to the boundary conditions is captured by a phase transition, termed
as TPT, as we interpolate by Hamiltonian deformation between different boundary conditions. We have invoked the notion of
parity conservation and anyonic symmetries and have established that a break in either one of the above symmetries is
sufficient to characterize the TPT. To further consolidate the presence of a TPT, we have numerically analyzed signatures
such as ground state degeneracy, TEE and have introduced the notion of open-loop operator whose expectation value captures the
phase transition. While the ground state degeneracy and expectation value of the open-loop operator provide an estimate
of the critical strength, we have re-established the fact that TEE remains constant and is thereby ineffective in
detecting the above introduced TPT's. 

  Having established the notion of TPT in a closed setup, we extend it to an open quantum setup. The phase transitions in
an open setting are associated with the steady states obtained by solving the LME. To this extent, we have introduced
collapse operators, whose dissipative rates are a function of the interpolation parameter $\lambda$. Due to the above
construction, the dynamics cool the product state into distinct TO steady states at different
$\lambda$, with the extremities being mapped to the relevant TC ground states, thereby encoding a TPT at some
critical $\lambda$. We have shown that the expectation value of the open-loop operator is still relevant and is effective in detecting such TPT's in an open setup.

  In this paper, having analyzed the presence of TPT's in various closed and open setups, it would be interesting
to gain an insight into the stability of topological order due to different boundaries, in a dynamical setting as the system is
quenched across a TPT \cite{Tsomokos2009}.  The introduced TPT's being characterized by non-local order parameter, it would be
interesting to study the notion of Kibble-Zurek like mechanism in both closed and open setting \cite{Chandran2013}. There has
been a recent proposal to define topological phases in the context of open quantum systems \cite{Coser2019}, it would be
interesting to study the TPT in an open setup introduced in this work with the above definition. Experimentally, there has been progress
in realizing the ground states of the TC Hamiltonian as in Ref.~\cite{Sameti2017}, which also includes open system 
scenarios with various noise protocols, it would be interesting to study the realization of proposed engineered collapse 
operators in such a setup. Also, there has been recent progress in preparing quantum states using variational quantum circuits \cite{Kokail2019}, 
it would be interesting to extend the above protocol to realize the interpolated topological steady states by including 
suitable variational dissipators. Some of the immediate extensions would be to detect the presence of similar TPT's in the 
context of other abelian and non-abelian models with an aim to develop other relevant signatures.

\begin{acknowledgements}
We are grateful to Hendrik Weimer and Javad Kazemi for helpful conversations and insightful comments. This
work was funded by the DFG within SFB 1227 (DQ-mat) and SPP 1929 (GiRyd). AB is supported by Research Initiation 
Grant (RIG/0300) provided by IIT-Gandhinagar.
\end{acknowledgements}

\vspace{0.5cm}

\textit{Note:} While preparing  this manuscript we became aware of the following work Ref.~\cite{Lichtman2020}.
The authors have discussed the case presented as in Sec.~\ref{tcbv} of the current work. 

\appendix

\section{Interpolating between mixed boundaries on either end}

\begin{figure}[h!]
\begin{center}
\includegraphics[width=0.6\linewidth]{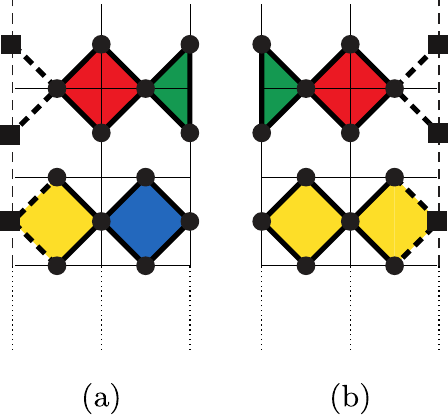}
\caption{\label{afig1}TC on a cylinder with mixed boundary conditions (a) rough boundary on the left and smooth
boundary on the right (b) smooth boundary on the left and rough boundary on the right.}
\end{center}
\end{figure}

We interpolate between TC on a cylinder with mixed boundary conditions as in Fig.~\ref{afig1} (we interpolate between (a)
and (b) as $\lambda$ is varied from 0 to 1). The TPT is characterized by the energy gap closing at $\lambda=0.5$ and
belongs to the class of $\Tilde{G}_{\lambda=0} =  \Tilde{G}_{\lambda=1}$. There is neither parity conservation, as
excitations can be singly drawn from the boundary, nor anyonic symmetry, due to the condensation properties at the
boundary, for all $\lambda$, implying that it is not necessary that every TPT is accompanied by a broken symmetry. In the main
discussion, we referred to the parity being broken with respect to $e, m$-type excitations
without laying much emphasis on the choice of the boundary of the cylinder i.e., left or right physical boundary.  We
observe that by specifying the parity symmetry with respect to a particular physical boundary, allows us to state the
following: Either a break in the parity with respect to a particular physical boundary or break in the anyonic symmetry is
necessary and sufficient to characterize the presence of a TPT.

Extending the above implication to the current scenario, it is evident that  that the parity of $e(m)$-type excitations is
preserved with respect to the right (left) physical boundary in the limit of $\lambda=0$, while is broken in the limit of
$\lambda=1$. Therefore, we have substantiated that imposing stronger conditions on the parity preservation leads to a
bi-implication between the presence of TPT and the parity conservation, anyonic symmetries. The above statement may be 
generalized for any abelian quantum doubles, as the parity of atleast one of the superselection sectors is broken due to the
condensation at the boundary.

\section{TPT's with the domain wall intact}
\begin{figure}[h!]
\begin{center}
\includegraphics[width=0.85\linewidth]{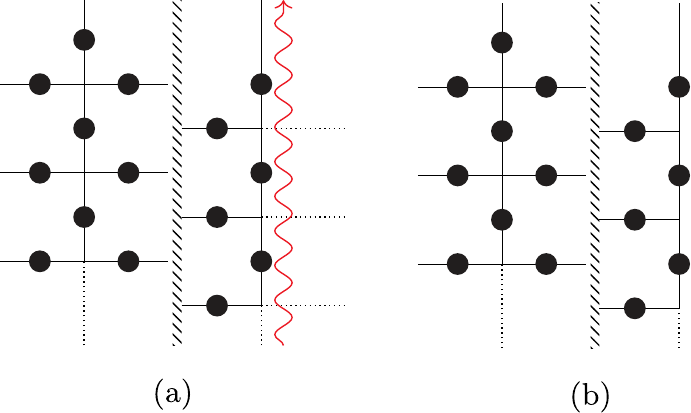}
\caption{\label{afig2}(a) TC on a torus with a domain wall. The red snake represents the interpolation cut which breaks the periodicity along some other rail other than the domain wall leading to (b) TC on a cylinder with mixed boundaries
on either end with the domain wall intact.}
\end{center}
\end{figure}

In every scenario discussed above, we have observed that the TPT is characterized by break in parity conservation
of either $e, m$ excitations or both due to the introduction of relevant boundary conditions. In this section, we present
a scenario where the TPT is solely characterized by the break in anyonic symmetry with no conservation in parity, at all
$\lambda$. To this extent, we consider the TC on a torus with domain wall ($\lambda=0$) and instead of interpolating along
the domain wall we cut through the periodic boundary as in Fig.~\ref{afig2} to a cylinder with mixed boundary
with the domain wall intact ($\lambda=1)$. The interpolation encodes a TPT as the GSD in the limit of $\lambda=0$ is 2
while in the limit of $\lambda=1$ is 4. 

In the limit of $\lambda=0$, there is no conservation in parity due to the presence of domain wall although the anyonic
symmetry is conserved. On the other hand at $\lambda=1$ it is still possible to draw single excitations from the boundary
thereby there is no conservation in parity while the anyonic symmetry is also broken due to the introduction of open
boundaries. Therefore, in this case the TPT is solely characterized by the break in anyonic symmetry.

\section{TPT's arising out of simultaneous dissection and gluing}
\begin{figure}[h!]
\begin{center}
\includegraphics[width=0.85\linewidth]{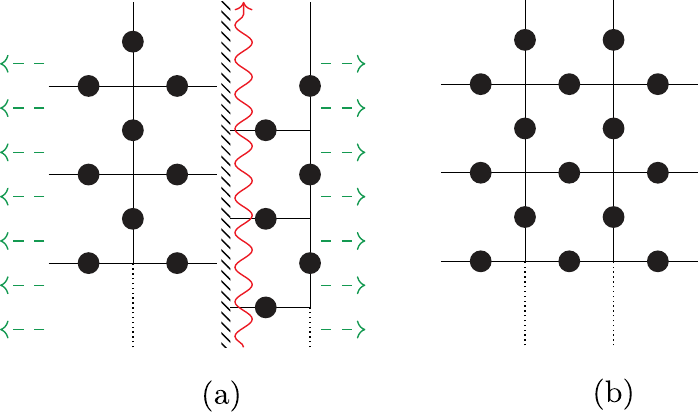}
\caption{\label{afig3}Interpolating via simultaneous dissection and gluing, the red snake represents the dissection while
the dashed green arrows represent the gluing action. (a) TC on a clylinder with mixed boundaries and a domain wall (b)
TC on cylinder with rough boundaries on either end.}
\end{center}
\end{figure}

In this section, we introduce a TPT arising out of simultaneous dissection and gluing along two different boundaries.
To this end, we consider the TC Hamiltonian on cylinder with mixed boundaries along with a domain wall in the limit of
$\lambda=0$, being mapped to TC Hamiltonian on a cylinder with a rough boundary at either end in the limit of $\lambda=1$,
see Fig.~\ref{afig3}. 

The TPT is marked by the change in GSD as it maps from 4 in the limit of $\lambda=0$ to 2 in the
limit of $\lambda=1$. Additionally, we also note that the parity conservation is preserved with respect to $m$-type
excitations in the limit of $\lambda=1$ while it remains broken in the limit of $\lambda=0$.

\section{Dissipative interpolation via imperfect cooling}
In Sec.~\ref{sec_idiss}, we have introduced collapse operators whose action leaves the state invariant in the absence of the
excitations or diffuse/annihilate the excitations when present. In this section, we introduce collapse operators as in 
Eq.~\ref{copa}, where the $A_{v}(B_{p})$ operators along the interpolation cut are additionally scaled by the relevant
interpolation parameter.

\begin{equation}\label{copa}
\begin{gathered}
\begin{split}
    c_{v}^{t}(\lambda) &= \frac{\sqrt{\gamma_{v}}}{2}(1-\lambda)\sigma_{z}^{(i)}(\mathds{1}-(1-\lambda)A_{v}^{\mathbin{
    \blacklozenge}}), \\
    c_{f}^{t}(\lambda) &= \frac{\sqrt{\gamma_{f}}}{2}(1-\lambda)\sigma_{x}^{(j)}(\mathds{1}-(1-\lambda)B_{f}^{\mathbin{
    \blacklozenge}}), \\
    c_{v}^{o}(\lambda) &= \frac{\sqrt{\gamma_{v}}}{2}\lambda\sigma_{z}^{(i)}(\mathds{1}-\lambda
    A_{v}^{\Rtri}), \\
    c_{f}^{o}(\lambda) &= \frac{\sqrt{\gamma_{f}}}{2}\lambda\sigma_{x}^{(j)}(\mathds{1}-\lambda
    B_{f}^{\Ltri}),
\end{split}
\end{gathered}
\end{equation}

\begin{figure}
\begin{center}
\includegraphics[width=\linewidth]{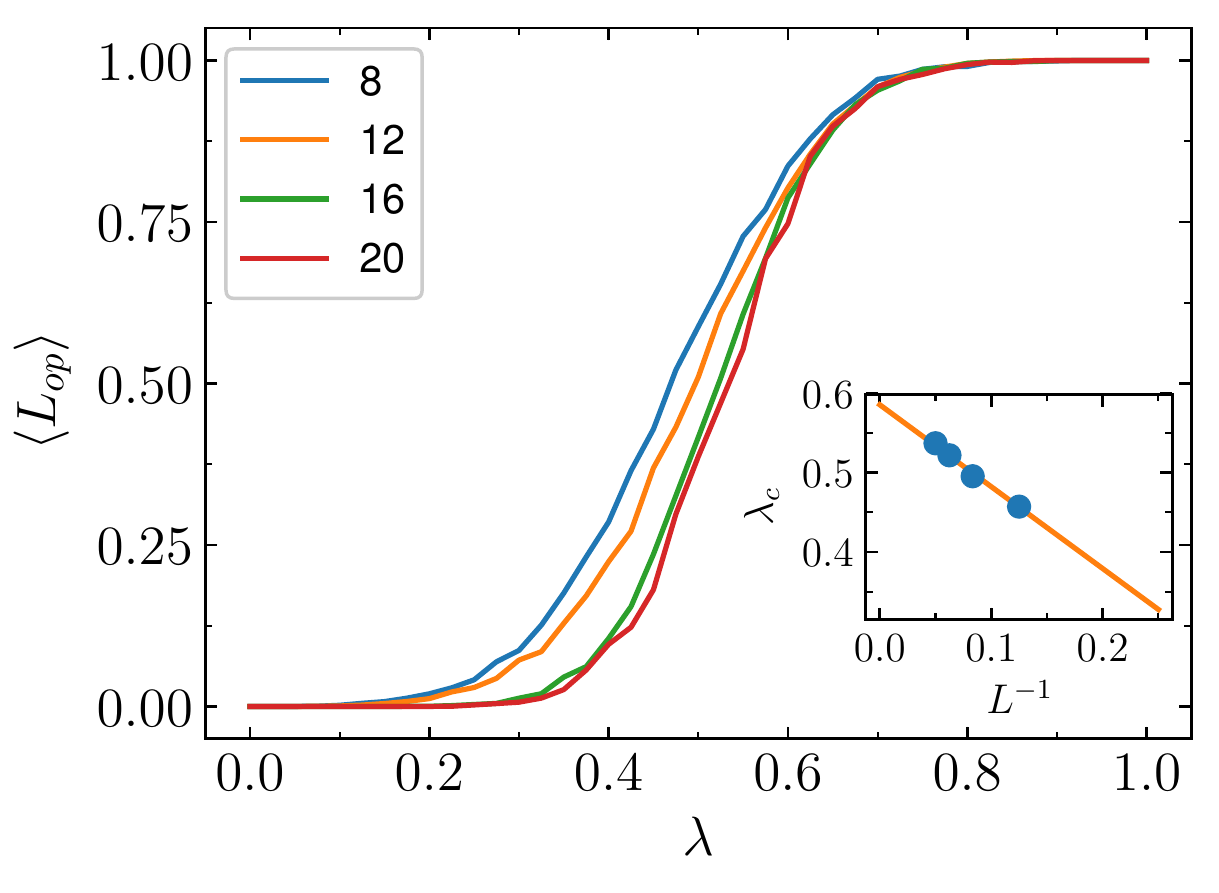}
\caption{\label{afig4}Expectation value of the longest open-loop operator with respect to the interpolation strength, 
$\lambda$ with the labels denoting the different system sizes. (Inset) Finite size scaling to obtain the critical strength,
$\lambda_{c}$.}
\end{center}
\end{figure}

The key difference between these collapse operators and the ones introduced earlier, as in Eq.~\ref{cop},
is given by the fact that in the absence of excitations, the former induces additional excitations while the latter leaves the
state invariant. To gain further insight into the phase transition, we compute the expectation value of the open-loop operator
with respect to the interpolation strength, $\lambda$, see Fig.~\ref{afig4}. By performing finite size
analysis, we obtain $\lambda_{c} = 0.586 \pm 0.001$ which is lower compared to the earlier case. 

\bibliographystyle{aip}
\bibliography{bib.bib}

\end{document}